\documentclass[twocolumn,aps,showpacs,prb]{revtex4}
\usepackage{graphicx}

\usepackage{amssymb}
\usepackage{amsmath}
\usepackage[latin1]{inputenc}
\usepackage{bm}

\newcommand{\nn}{\nonumber}
\newcommand{\so}{\mathrm{so}}
\newcommand{\uu}[1]{\bm{{#1}}}
\newcommand{\vc}[1]{\bm{{#1}}}
\newcommand{\vco}[1]{\hat{\bm{#1}}}
\newcommand{\Tr}{\mathrm{Tr}}

\newcommand{\be}{\begin{equation}}
\newcommand{\ee}{\end{equation}}
\newcommand{\bea}{\begin{eqnarray}}
\newcommand{\eea}{\end{eqnarray}}

\begin{document}

\title{Geometric phases in semiconductor spin qubits: Manipulations and decoherence}
\author{Pablo San-Jose$^{1}$, Burkhard Scharfenberger$^{1}$, Gerd Sch\" on$^{1}$,
Alexander Shnirman$^{1,2}$, and Gergely Zarand$^{3}$}

\affiliation{ $^1$ Institut f\"{u}r Theoretische
Festk\"{o}rperphysik and DFG-Center
    for Functional Nanostructures (CFN), Universit\"{a}t Karlsruhe,
    D-76128 Karlsruhe, Germany\\
$^2$ Institut
f\"{u}r Theoretische Physik, Universit\"{a}t Innsbruck, A-6020
Innsbruck, Austria\\
$^3$ Institute of Physics, Technical University Budapest,
Budapest, H-1521, Hungary}
\date{\today}

\begin{abstract}
We describe the effect of geometric phases induced by
either classical or quantum electric fields acting on
single electron spins in quantum dots in the presence of
spin-orbit coupling. On one hand, applied electric fields
can be used to control the geometric phases, which allows
performing quantum coherent spin manipulations without
using high-frequency magnetic fields. On the other hand,
fluctuating fields induce random geometric phases that lead
to spin relaxation and dephasing, thus limiting the use of
such spins as qubits. We estimate the decay rates due to
piezoelectric phonons and conduction electrons in the circuit,
both representing dominant electric noise sources with
characteristically differing power spectra.
\end{abstract}

\maketitle

\section{Introduction \label{sec:introduction}}

Recent demonstrations of coherent single-electron spin
control and measurement in semiconductor quantum
dots~\cite{Elzerman04,Petta05,Koppens06} represent
milestones on the way to quantum-state engineering with
spin qubits~\cite{Loss98}. In addition, work on  coherent
spin transport in
nanostructures~\cite{Kato05,Hankiewicz06,Engel05,Awschalom07}
has revealed new possibilities for next-generation
spintronic devices. The key behind these emerging
technologies is the long spin coherence time in
semiconductor materials.

The standard technique for addressing and manipulating
spins in semiconductors is electron spin resonance (ESR)
controlled by external ac magnetic
fields~\cite{Koppens06}. Alternatively,
effective internal magnetic fields can be generated via the spin-orbit (SO) interaction by applied electric fields.
Proposals for coherent control of confined electron spins
based on this combination have
been put forward~\cite{Duckheim06,Bulaev06, Golovach06, Stano06b, Tang06}, and  experimental progress has been reported~\cite{Stotz04,Kato04}.
At the same time, SO interaction makes the electron spin sensitive to the electric noise ubiquitous in typical solid state
environments~\cite{Abrahams57,Dyakonov72}.
The combination of both provides an important
mechanisms by which electron spins  decay and lose
coherence~\cite{Khaetskii00,Khaetskii01,Woods02,Golovach04,Sherman05,sanjose06}.
Such electric field fluctuations are generated, e.g., by lattice vibrations, but in low magnetic fields the Nyquist noise in the electrodes may be
dominant~\cite{sanjose06}.

The details of the spin-orbit mediated interaction between the
electric field and the electron spin hide some interesting
twists. As we discuss below, in suitable time-dependent
electric fields the spin of an electron confined in a
quantum dot acquires a {\em non-Abelian
geometric phase} (a generalized Berry phase).
On the one hand, this geometric quantum state evolution allows for new spin
manipulation strategies purely controlled by electric fields. In comparison to the alternative, ESR manipulation by ac magnetic fields,
geometric spin manipulation is potentially more robust, since it is
not affected by gate timing errors and certain control voltage inaccuracies.

\begin{figure}
\includegraphics[width=7 cm]{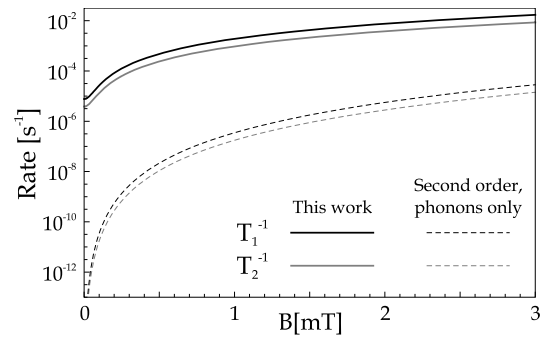}
\caption{Spin relaxation and decoherence rates in semiconducting quantum dots.  Results of the present work (solid lines) are compared to those
of Ref. \onlinecite{Golovach04} (dashed lines),
where the effect of piezoelectric phonons is analyzed to second order in the coupling to piezoelectric phonons.
We find that geometric dephasing (evaluated to fourth order) leads to a saturation at low magnetic fields.
Deviations between both results at intermediate field values are mainly due to the Nyquist noise of the conduction electrons.
At higher magnetic fields (not shown) the results coincide.
Parameters correspond to a parabolic quantum dot in GaAs with $\omega_0=5$K at $T=100$mK.
\label{fig:ratelinearlog}}
\end{figure}

On the other hand, in fluctuating electric fields the accumulation of random geometric phases leads to spin relaxation and decoherence for low and
even  vanishing magnetic fields~\cite{sanjose06}, predominantly induced by  high-frequency noise, $\hbar \omega\sim k_{\rm B} T \gg \mu_B B$.
This saturation of the spin decay rate at low fields has been overlooked in the literature~\cite{Khaetskii01,Stano06,Semenov06}, although
a similar connection had been discussed for free electrons in the presence of disorder scattering~\cite{Serebrennikov04}.
In comparison to the previously discussed spin-orbit mediated mechanisms~\cite{Khaetskii00,Stano05} the geometric spin decay is of higher order in the electric field. It requires a minimum of two independent noise sources coupled to two non-commuting components of the electron spin, whereby the non-Abelian character of spin rotations (properties of the SU(2) group) becomes relevant.

The present paper is devoted to a detailed study of the
spin evolution of a confined electron in a time-dependent electromagnetic
environment. It extends earlier work~\cite{sanjose06} which is based on a perturbative approach that allows handling quadrupolar and
octupolar fluctuations but is restricted to small and adiabatic electric
fields. Here we develop alternative adiabatic methods, that
allow describing large electric fields and corresponding displacements.

To begin with, we consider in Sec.~\ref{sec:geomphases} a
``semi-classical'' electron moving along a fixed trajectory.
This limit describes the situation where an electron is confined in a narrow quantum dot potential which is shifted by applied classical electric fields.
The spin-orbit interaction, which is treated in leading order in the small parameter $x_0/l_\so$ (dot size over SO length), induces a geometric spin precession tied to the electron's motion.
We illustrate how electric fields can be used
to manipulate the spin state via the SO interaction. We further provide
a qualitative picture for geometric dephasing (Sec. \ref{sec:Berryrel} and \ref{sec:clafluc}), if the fields are classical fluctuating fields completely characterized by their power spectrum.
A more rigorous analysis based on a fully quantum-mechanical treatment of the electromagnetic field follows in later sections.

Next, in Sec.~\ref{sec:QeCf}, we treat
the confined electron fully quantum-mechanically taking into account the  spin-texture of the confined orbitals in a (parabolic) quantum dot.
This approach allows us to study larger dots with intermediate and high values of
$x_0/l_\so$. Spin-dressing effects in larger dots lead to a renormalization of the g-factor  that also influences the geometric spin evolution. We
first consider the effect of a {\sl classical} applied field and then
generalize to the case where also the electromagnetic
field is treated quantum-mechanically within a path integral approach.
This method, presented in Sec.~\ref{sec:QeQf},  allows us to compute
 the dephasing and relaxation rates of the electron spin
by means of a systematic diagrammatic perturbation theory.

In section~\ref{sec:results} we present our results for the
spin relaxation and decoherence rates in semiconducting quantum dots.
We find that at low temperatures and low magnetic fields, $B<1$T, {\em Ohmic fluctuations} originating from the electrodes dominate over the phonon effects in their infuence on the spin relaxation~\cite{sanjose06}. This is simply due to the fact that the spectrum of excitations of a metallic electrode is much denser than that of phonons at low energies. At still lower magnetic fields and temperatures, however, the dominant mechanism leading to relaxation is the accumulation of random geometric phases. This leads to a saturation of the decay rates at $B=0$. Both features are
included in the data shown in Fig.~\ref{fig:ratelinearlog} and compared to previous predictions. Further results including the
 temperature dependence under typical experimental conditions
are presented in section~\ref{sec:results}.

Finally, in Sec.~\ref{sec:manipulation}, we study an electron in an array of quantum dots and show that in the presence of SO interaction arbitrary spin rotations can be reached by a series of coherent electron tunneling processes in a multi-dot geometry. We discuss the possibility to use such multi-dot systems for geometric electron spin manipulations.

\section{Geometrical spin evolution in a semiclassical picture
\label{sec:concepts}}

\subsection{Geometric phases and spin rotations \label{sec:geomphases}}

Geometrical phases, introduced in the pioneering work of Berry~\cite{Berry84}, describe how a quantum system in a non-degenerate state
evolves when
adiabatically driven around a closed loop in a control
parameter space.  Apart from  a  dynamical phase,
depending on the energy and elapsed time, the state acquires  a
contribution, called the Berry phase, that only depends on
the geometry of the loop in parameter space. An important
extension of the Berry phase covers the case when the initial state belongs to
a {\em degenerate} subspace~\cite{Wilczek84}. Then the
geometric phase is replaced by a non-Abelian unitary
transformation of the initial state within the degenerate
subspace, and again, this unitary transformation depends only on the
geometry of the loop.

This non-Abelian generalization of the Berry phase is of
direct relevance for an electron confined in a quantum dot:
Time reversal symmetry in the absence of a magnetic field implies that
the spectrum of a quantum dot is a collection of
time-reversed Kramers doublets even in the presence
of spin-orbit interaction.
In the absence of SO coupling these states correspond to pure spin states.
In the general case the doublets persist, and we call them pseudospin states.

The adiabatic variation of  the position (and possibly
shape) of the  potential confining the electron
is described by a time-dependent Hamiltonian $H(t)$.
Adiabaticity ensures that an electron which at time $t=0$ is in the ground state doublet remains within the corresponding subspace of
instantaneous ground states of $H(t)$.
The adiabatic time evolution thus appears as a non-Abelian
SU(2) transformation within the ground state doublet~\cite{Wilczek84}.
In the presence of spin-orbit coupling the unitary transformation results in a geometric spin rotation, which only depends
on the trajectory of the electron in real space.

To illustrate the geometric spin rotation in more detail
we consider a 2DEG, assumed to be grown along the $[001]$ direction.
An electron in the 2DEG experiences a spin-orbit coupling which takes the form ($\hbar=1$, $x$ axis along the $[100]$ direction)
\be
\mathcal{H}_\so=\alpha\left(\hat{p}_y\hat{\sigma}_x-\hat{p}_x\hat{\sigma}_y\right)+\beta\left(\hat{p}_y\hat{\sigma}_y-\hat{p}_x\hat{\sigma}_x\right)
=\frac{1}{m}\vco{p}\uu{\lambda}^{-1}_\so\vco{\sigma}\;.
\label{Hsodef}
\ee
Here $\vco{p}$ denotes the momentum of the electron in the $xy$ plane of
the 2DEG and  $\vco{\sigma}/2$ is the spin operator.
The Rashba ($\alpha$) and
linear Dresselhaus ($\beta$) couplings can be lumped into the
spin-orbit tensor $\uu{\lambda}_\so$
\begin{equation}
\uu{\lambda}_\so^{-1}\equiv m\left(\begin{array}{cc} -\beta&-\alpha\\
\alpha&\beta\end{array}\right).
\end{equation}
An implicit summation  over the $x,y$ coordinates is assumed, i.e.\
$\vco{p}\uu{\lambda}^{-1}_\so\vco{\sigma}
=\sum_{\mu\nu=x,y}\hat{p}_\mu\left({\lambda^{-1}_\so}\right)_{\mu\nu}\hat{\sigma}_\nu$.
The tensor $\uu\lambda_\so$ sets the scale for the spin-orbit length
$l_\so\equiv\sqrt{|\det\uu\lambda_\so|}=(m\sqrt{|\alpha^2-\beta^2|})^{-1}$,
that defines  the  typical length scale of  spin textures.
In typical GaAs/AlGaAs semiconductor heterostructures
we have $l_\so\approx 1-5\mathrm{\mu m}$.\cite{Amasha06}

We further assume that by lateral structuring of the 2DEG
a parabolic quantum dot, $V(\vco r)=\frac{1}{2}m\omega_0^2\vco{r}^2$,
is created with energy scale $\omega_0$ related to the orbital size $x_0=1/\sqrt{m\omega_0}$. Typical dot sizes are in the range $x_0\approx 30-100 \mathrm{nm}$. Hence $x_0/l_\so$ is usually small, of the order of $0.1-0.01$.
The electron is assumed to be strongly confined
in the ground state of the potential.
By applying electric fields (modifying the confining potential) we move the dot and the electron along a trajectory $\vc R_\mathcal{C}(t)$.

In the considered limit, $x_0/l_\so \ll 1$,
the degenerate doublet of states of the confined electron simply differ in the spin orientation. Eq. (\ref{Hsodef}) then implies that
the confined spin experiences  an effective magnetic field,
 $H_\so=\vc B_\so(t) \cdot\vco\sigma$, with  $\vc B_\so=\frac{1}{m} \langle\vco
p\rangle \uu\lambda_\so^{-1}$.
For a strongly confining potential we have
$\langle \vco p\rangle \approx  m\vc{\dot R}_\mathcal{C}$.
Therefore the induced spin-evolution is described by~\cite{Coish06}
\begin{eqnarray}
\nonumber
U_\mathrm{ad}&=&{\rm T}\;\exp{\left(-i\int_0^tdt\;\vc{\dot R}_\mathcal{C}\uu\lambda_\so^{-1} \vco\sigma\right)}\\
&=&
{\rm P}\; \exp{\left(-i\int_\mathcal{C}  d\vc{R}_\mathcal{C}
\uu\lambda_\so^{-1}\; \vco\sigma\right)}
\label{Uadclassic}
\end{eqnarray}
with $\rm T$ and $\rm P$ denoting time- and path-ordering
operators, respectively. The tag `ad' refers to the constraint of
adiabatically slow paths, $|\vc{\dot R}_\mathcal{C}|\ll x_0\omega_0$,
 which guarantees that
the evolution affects the electron spin but does not induce
transitions to excited dot states.
Clearly, $U_\mathrm{ad}$
depends only on the {\em geometry} of a given path  $\mathcal{C}$
itself, not on the time dependence of $\vc R_\mathcal{C}$.
As we shall show in Sec.~\ref{sec:QeCf}, this  result
can be generalized to larger parabolic dots for which
spin texture effects are important. The main difference in that case
is that $\vc \;\uu\lambda_\so^{-1}$ is replaced by a renormalized tensor ${\vc \;\uu{\tilde \lambda}_\so^{-1}}$.

\begin{figure}
\includegraphics[width=8.5 cm]{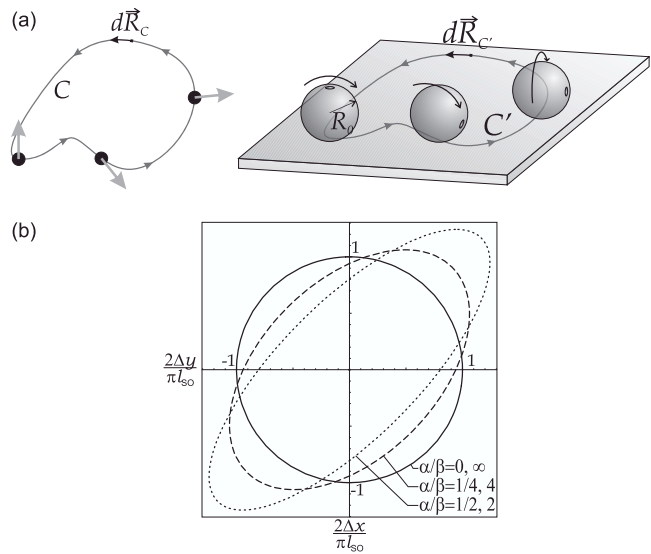}
\caption{(a) The geometric spin precession due to SO interaction for an
electron adiabatically moving within a 2DEG
(left) is equivalent to the changing orientation of a sphere rolling
on a plane (right). Both paths are related in a one-to-one fashion to each other. (b) The length of a straight path required to perform a spin
flip versus the angle of the path relative to a crystal axis
for several ratios $\beta/\alpha$ with a constant $l_\so$. The area of
the ellipses remains constant and equal to $\pi
l_\so^2$. \label{fig:rolling}}
\end{figure}

In order to visualize the spin rotation described by Eq.~\eqref{Uadclassic} it is instructive to use the connection between SU(2) and
SO(3) rotations.
 The change of the orientation of a sphere rolling
on a plane along a path $\vc{R}_\mathcal{C'}$ is  characterized by
a rotation operator
\begin{eqnarray}
U_\mathrm{sph}&=&P\exp\left(-i\int_{C'}d\vc
R_\mathcal{C'}\, \uu\lambda_\mathrm{sph}^{-1}\vco A\right)
\label{Usph}\\
\uu\lambda_\mathrm{sph}^{-1}&=&\frac{1}{R_0}\left(\begin{array}{cc}
0&1\\-1&0\end{array}\right),
\end{eqnarray}
where the components of $\vco A$ denote the standard SO(3) generators.
Comparing (\ref{Uadclassic}) and (\ref{Usph}) and
choosing the radius of the sphere to be  $R_0\equiv l_\so/2$
we note that a spin rotation
$U_{\rm ad}$ associated with a path $\vc{R}_\mathcal{C}$
maps onto a rotation of the sphere rolling along
a trajectory  $\vc{R}_\mathcal{C'}(t)\equiv
2\vc{R}_\mathcal{C}(t) \uu\lambda_\mathrm{sph}$
(see Fig.~\ref{fig:rolling}).
If only Rashba coupling is present
then the paths $C$ and $C'$ have the same shape. In general, if also Dresselhaus coupling is present, then $C$ and $C'$ have different shapes, but the qualitative analogy persists.

In a 2DEG,  with both Rashba and Dresselhaus couplings present, the change of the spin orientation induced by moving
the confined electron  along a straight line depends on the direction
of trajectory relative to a crystal axis.
The distance to perform a spin-flip is strongly anisotropic for
$\alpha\sim \beta$,  being enhanced in the $[110]$  direction.
This property is illustrated in Fig.~\ref{fig:rolling}(b).

\subsection{Spin manipulation and spin relaxation \label{sec:Berryrel}}

To illustrate how the geometric phase evolution given by  Eq.~\eqref{Uadclassic} can be used to manipulate the spin state we consider a quantum dot orbiting $N$ times around a closed circular path, $\vc R(t)=R_0(\cos\omega
t,\sin\omega t,0)$,  with period $t_0=2\pi/\omega$.
The spin-orbit length is assumed much larger than the loop size,
$l_\so\gg R_0$. For simplicity we assume pure Dresselhaus coupling. The resulting spin precession is then described by $U_\mathrm{ad} = [U_\mathrm{ad}(t_0)]^N$,
where the evolution operator of  a complete loop is
$$
U_\mathrm{ad}(t_0)={\rm P}\;\exp\left[-i\int_0^{2\pi}
d\phi\;\frac{R_0}{l_\so}\left(\hat{\sigma}_x\sin\phi +
\hat{\sigma}_y\cos\phi\right)\right]\;.
$$
To second order in $R_0/l_\so$ it reduces to
\begin{eqnarray}
U_\mathrm{ad}(t_0) &\approx&
1-i\sigma_z \frac{2\pi R_0^2}{l_\so^2}\approx
\exp\left(-\frac{i}{2}\frac{4
A(t_0)}{l_\so^2}\hat{\sigma}_z\right)\;,
\label{Uad00}
\end{eqnarray}
where $A(t_0)=\pi R_0^2$ denotes the area enclosed by a loop.
As will be shown in Sec.~\ref{sec:QeCf},
Eq.~\eqref{Uad00} is valid  for small closed loops of arbitrary shape.
The full time evolution after $N$ circles can be approximated as
\be
U_\mathrm{ad} \approx\exp\left(-\frac{i}{2}\frac{4
A(t)}{l_\so^2}\hat{\sigma}_z\right) \; ,
\label{Uad0}
\ee
where $A(t)=N\;A(t_0)$ is the area swept by $N=t/t_0$ repetitions
of the loop. Obviously the motion in small closed loops
corresponds to an effective static magnetic field,
$B_{\rm eff} = 4 A(t_0)/(t_0\;l_\so^2)$ pointing along the
$\hat z$-direction.

Eq.~\eqref{Uad0} also holds if the trajectory $\vc R_\mathcal{C}(t)$ is a stochastic variable, typically induced by a fluctuating
 electric field, $\vc R_\mathcal{C}(t) \propto \vc E$.
In this case the area $A(t)$ in
Eq.~(\ref{Uad0}) is a random variable whose dispersion increases linearly in time,   $\langle
A(t)^2 \rangle\approx A_0\; t$. Assuming that
$A(t)$ is Gaussian distributed, we conclude that  off-diagonal
elements of the spin density matrix decay as
\begin{eqnarray}
\label{dephasing} \langle \hat{\sigma}_x\rangle \sim \langle \hat{\sigma}_y\rangle \sim
\exp\left(-8\frac{\langle A(t)^2\rangle}{l_\so^4}\right)\ .
\end{eqnarray}
Thus the accumulation of - random - geometric phases leads to dephasing, even in the absence of external magnetic fields.
 The coefficient $A_0$ introduced above depends on
the amplitude and frequency of typical electromagnetic fluctuations, and its proper computation requires a fully quantum-mechanical treatment of the
electromagnetic field~\cite{sanjose06}, reviewed in Section~\ref{sec:QeQf}.
At this point we note that the area itself is proportional to the
square of the electric field, $E_\omega^2$, with $\omega\sim T$ the
typical frequency of electromagnetic
fluctuations, and the characteristic time for making a closed loop
is  $\sim 1/\omega$.   Hence, the geometrical relaxation rate is
proportional to $\langle \omega\;|E_\omega|^4\rangle$. E.g., for Ohmic
fluctuations we have $|E_\omega|^2\sim T^2$ and consequently the corresponding relaxation rate scales as $\sim T^5$. Note that this geometric dephasing is of fourth order in $E$ and, therefore, requires fourth order perturbation
theory in the electromagnetic field.

\subsection{Spin relaxation in a magnetic field
\label{sec:clafluc}}

So far we assumed that no external magnetic field is applied.
Now we study the relaxation and dephasing of a spin in an
in-plane magnetic field $\vc B$ in the presence of
 a classical stochastic field,
$\vc B_\so=\vc{\dot R}_\mathcal{C}\uu\lambda_\so^{-1}=(B_{\so x},B_{\so
y}, B_{\so z})$, characterized by the power spectrum
\begin{equation}
S_{ij}(\omega)\equiv \int_{-\infty}^\infty dt e^{i\omega t} \langle
B_{\so i}(t)B_{\so j}(0)\rangle .
\end{equation}
The external field  $\vc B$ induces a precession of the spin around its
direction  with   Larmor frequency $\omega_B=-g\mu_B|\vc B|$, while the
fluctuating field $\vc B_\so$ leads to relaxation and dephasing
with rates  $\Gamma_1 \equiv T_1^{-1}$ and $\Gamma_2 \equiv
T_2^{-1}$, given to leading order by the standard
expressions~\cite{Bloch57,Leggett87, Weiss99}
\bea
\label{eq:T_1_BR}
\Gamma_1^{(2)}& = &2\,S_\perp(\omega_B) \ ,
\\
\label{eq:T_2_BR}
\Gamma_2^{(2)} &= &\frac{1}{2}\,\Gamma_1^{(2)} +
2\,S_{\parallel}(0)\ .
\eea
Here $S_\perp(\omega)$ and $S_\parallel(\omega)$
are the spectral densities of the components of $\vc B_\so$
perpendicular and parallel to $\vc B$, and the superscript $(2)$, introduced to distinguish from later extensions,
refers to second order. In case of isotropic random motion of
the dot and purely Rashba (or purely Dresselhaus) spin-orbit
coupling we have $S_\perp(\omega) = S_\parallel(\omega)=S(\omega)$.
Typically $S(\omega=0)=0$ if the electron remains localized in space,
since only unbounded paths can produce a static $\vc B_\so\sim\vc{\dot R}_\mathcal{C}$.
Therefore the last term of Eq.~(\ref{eq:T_2_BR}) vanishes in this
case and we obtain  in leading order~\cite{Golovach04} $T_2 = 2 T_1$.

We can also estimate the relaxation rate induced by the
Berry phase mechanism, which is of higher order than
 expressions Eqs.~(\ref{eq:T_1_BR}) and (\ref{eq:T_2_BR}):
The effective magnetic field generated
by the Berry phase term is proportional to
$B_{\rm eff}\sim \dot A \sim \vc{\dot R}_\mathcal{C}\times
\vc{R}_\mathcal{C}$, and is perpendicular to the in-plane
magnetic field. The relaxation rate of the spin to fourth order in the coupling to the stochastic field is proportional
to the Fourier component of the autocorrelation function
of this effective  field at frequency $\omega\sim \omega_B$,
\begin{equation}
\Gamma_1^{(4)}
\propto \int d\omega (\omega^2 +\omega_B^2)\,
C_{\perp}\left(\frac{\omega_B+\omega}{2}\right)
C_{\parallel}\left(\frac{\omega_B-\omega}{2}\right)\;,
\label{eq:4order_estimate}
\end{equation}
where $C_{\parallel}$ and $C_{\perp}$ refer to the
spectral functions of the components $\vc{R}_{\mathcal{C},\parallel}$
and $\vc{R}_{\mathcal{C},\perp}$, parallel and perpendicular to the
in-plane magnetic field.

Strictly speaking, Eqs.~(\ref{eq:T_1_BR}-\ref{eq:4order_estimate})
hold only in the limit $\omega_B\gg \Gamma_1,\Gamma_2$. In the opposite
case,  $\omega_B < \Gamma_1,\Gamma_2$, also referred to as the
Zeno regime, the relaxation times $T_1$ and $T_2$
lose their meaning, and other dissipative rates need to be considered.
Although $\Gamma_1^{(2)}$ and $\Gamma_2^{(2)}$ vanish fast in the limit
$\omega_B\to 0$, $\Gamma_1^{(4)}$ {\em saturates} and scales to a
finite value. This is because the Berry phase relaxation
can also be induced by independent fluctuations of frequency
$\omega\sim T\gg \omega_B$, that have a 'beating' (frequency mismatch) at frequency
$\omega_B$, which is in resonance with the Larmor spin
precession.
As a result, for sufficiently small  $\omega_B$
but finite $T$ one always ends up  in the Zeno regime, dominated by the
Berry phase term,  Eq.~\eqref{eq:4order_estimate}.

\section{Beyond the semiclassical approach}

In this section we provide a fully quantum-mechanical treatment of a
confined electron's spin. We do this in two  steps:
First, we describe quantum-mechanically the spin and orbital state of the confined electron  in a classical time-dependent electric field. Then, we generalize this within a path integral formalism to include quantum-mechanical fluctuations of the electric field.

\subsection{Adiabatic approach in a classical electric field \label{sec:QeCf}}

\subsubsection{Motion in a parabolic dot}
\label{subsub:parabolic}

To be specific,  let us first consider an electron
in a parabolic confining
potential $V(\vco r)=\frac{1}{2}m\omega_0^2\vco{r}^2$ with
energy scale $\omega_0$ corresponding to a typical orbital
size $x_0=1/\sqrt{m\omega_0}$. For the moment we will
assume that the magnetic field is zero. In the presence of
an applied or fluctuating classical in-plane electric field
$\vc E(t)$ the Hamiltonian takes the form
\begin{eqnarray}
\mathcal{H}(t)&=&\frac{\vco{p}^2}{2m}+V(\vco{r})+e\vc
E(t)\, \vco r+\mathcal{H}_\so
\nonumber\\
&=&\frac{\vco{p}^2}{2m}+V(\vco{r}-\vc{R}_\mathcal{C}(t))
+\mathcal{H}_\so + {\cal C}(t)\;,  \label{H}
\end{eqnarray}
with ${\cal C}(t)\sim\vc E^2(t) $ being a time-dependent constant
of no relevance, and  $\vc{R}_\mathcal{C}(t)\equiv -e\vc E(t)/m\omega_0^2$.
Thus the effect of the homogeneous electric field is to move the center of
the potential along a  trajectory, $\vc{R}_\mathcal{C}(t)$. We assume that $\vc{R}_\mathcal{C}(0)=0$. Then
the Hamiltonians for times $t>0$
are related by the displacement
operator $\mathcal{W}(t)\equiv e^{-i\vco{p}\vc{R}_\mathcal{C}(t)}$,
\begin{equation}
\mathcal{H}(t)=\mathcal{W}(t)\mathcal{H}(0)\mathcal{W}^+(t) + {\cal C}(t).
\label{eq:unitary_transf}
\end{equation}
The center of the potential, $\vc{R}_\mathcal{C}(t)$, is
our external control parameter in the language of geometric
phases.
Eq.~\eqref{eq:unitary_transf}
 allows us to construct the instantaneous eigenstates
$\{|\tau_n(t)\rangle\}$  of $\mathcal{H}(t)$
 from the eigenstates $\{|\tau_n\rangle\}$ for  $\vc{R}_\mathcal{C}=0$,
 as
$\{|\tau_n(t)\rangle\}=\{\mathcal{W}(t)|\tau_n\rangle\}$.
The states come in Kramers doublets with $n=0,1,2,\dots$ and
$\tau=\uparrow,\downarrow$ denoting their
pseudospin orientation. We now introduce the 'adiabatic states' or 'center of motion
states' as
\be
|\tilde \Psi(t)\rangle \equiv  \mathcal{W}^+(t)| \Psi(t)\rangle\;.
\ee
Here the state $| \Psi(t)\rangle$ satisfies the time-dependent Schr\" odinger
equation with the Hamiltonian $\mathcal{H}(t)$, while the adiabatic states
$|\tilde \Psi(t)\rangle$ move
together with the dot and  satisfy the equation of motion
\be
i\partial_t |\tilde \Psi(t)\rangle =
(i \mathcal{\dot W}^+(t) \mathcal{W}(t) + \mathcal{H}(0)+ {\cal C}(t))
|\tilde \Psi(t)\rangle \;.
\ee
Using the explicit form of the operator $\mathcal W$ we thus find,
apart from a trivial overall phase generated by $ {\cal C}(t)$, that the evolution
of $|\tilde \Psi(t)\rangle $ is described by the effective Hamiltonian
\begin{eqnarray}
{\tilde H}_\mathrm{eff} =
\mathcal{H}(0) -\dot{\vc{R}}_\mathcal{C}(t)\;\vco{p}\, ,\label{Heff_unprojected}
\end{eqnarray}
and the corresponding evolution operator,
\begin{eqnarray}
\tilde U(t) = {\rm T}\exp\left\{-i\int_0^t dt'[\mathcal{H}(0)-\frac {d{\vc
R}_\mathcal{C}}{dt'}\cdot\vco p]\right\}\;.
\label{U}
\end{eqnarray}
 Clearly, from the above equations
it follows that if at time $t=0$ the electron occupies the ground state
 Kramers doublet, i.e.,
 $|\tilde \Psi(t=0)\rangle= \sum_\tau \alpha_\tau(t=0)|\tau_0\rangle$, and
 the external perturbation changes sufficiently slowly in time, $
|\dot{\vc{R}}_\mathcal{C}(t)|\ll x_0\omega_0$, then the
 second term in Eq.~\eqref{Heff_unprojected} cannot generate transitions to
 the excited states of $\mathcal{H}(0)$, and
$|\tilde \Psi(t)\rangle$ stays within the ground state doublet,
 $|\tilde \Psi(t)\rangle\approx  \sum_\tau \alpha_\tau(t)|\tau_0\rangle$.
Under these conditions the effective Hamiltonian can be approximated as
\begin{eqnarray}
{\tilde H}_\mathrm{eff}\approx -\dot{\vc{R}}_\mathcal{C}(t)\hat{P}_0\vco{p}\hat{P}_0
\, ,\label{Heff}
\end{eqnarray}
where $\hat{P}_0=\sum_\tau |\tau_0\rangle\langle \tau_0|$
 is the projector to the ground doublet of ${\mathcal H}(0)$,
and we took the energy of the ground state to be $E_0\equiv 0$.
The corresponding adiabatic evolution
operator within the center of motion ground state subspace then reads
\begin{eqnarray}
{\tilde U}_\mathrm{ad}(t)={\rm T}
\exp{\left(i\int_0^t dt' \dot{\vc{R}}_\mathcal{C}(t)\hat{P}_0\vco{p}\hat{P}_0\right)}
\label{Uad}\;.
\end{eqnarray}

\begin{figure}
\includegraphics[width=8 cm]{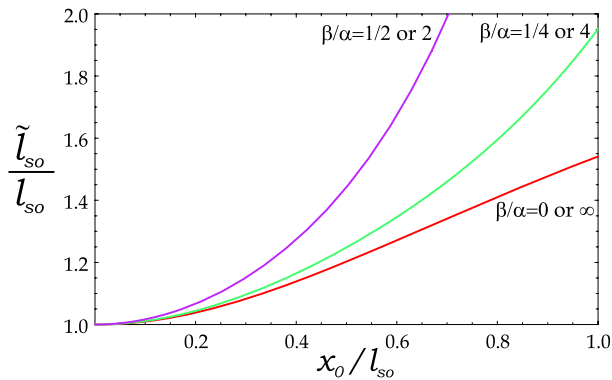}
\caption{(Color online) The dressed spin-orbit length
$\tilde l_\so$ as a function of the typical orbital size
$x_0=\sqrt{1/m\omega_0}$ for different ratios
$\beta/\alpha$. Exchanging $\alpha$ and $\beta$ yields
identical curves \label{fig:dressed}}
\end{figure}

Although not obvious, the
spin-orbit coupling plays an essential role in
  Eqs. (\ref{Heff}) and (\ref{Uad}).
As shown in Appendix~\ref{ap:Dressing} we have,
\begin{equation}
P_0 \vco p P_0=-\uu{\lambda}_\so^{-1}P_0 \vco \sigma P_0
=\uu{\lambda}_\so^{-1}\uu z\vco\tau,
\label{eq:renorm}
\end{equation}
where we introduced the
non-trivial $2\times 2$ spin-dressing tensor~\footnote{In general one
would expect $\uu z$ to be $3\times 3$ matrix, but due to reflection
symmetry $z\leftrightarrow-z$ of the
Hamiltonian, there is no coupling between the in-plane and
$z$ components of the spin and pseudospin, and hence only
the in-plane $2\times 2$ block of $\uu z$ is relevant.} $\uu z$,
that relates the matrix elements of the spin operator to
the pseudospin operator $\vco\tau$.
Thus an  adiabatic motion induces a pseudospin rotation
within the ground state multiplet. Note that the
formulas above are not perturbative in the spin-orbit coupling, and
they hold for parabolic dots of any size compared to the spin-orbit
length as long as the external field fluctuations are adiabatic.
Thus the effects of strong spin-orbit coupling can be fully taken into account by replacing $\uu{\lambda}_\so^{-1}$
by the dressed spin-orbit tensor
$\uu{\tilde\lambda}_\so^{-1}\equiv\uu\lambda_\so^{-1}\uu
z$, so that the adiabatic evolution operator  becomes
\begin{equation}
\tilde U_\mathrm{ad}= {\rm P}\exp\left(-i\int_C
d\vc{R}_\mathcal{C}\uu{\tilde{\lambda}}_\so^{-1}\vco\tau\right)
\,,\label{UadL}
\end{equation}
with  ${\rm P}$ being again the path ordering operator.
From Eq.~\eqref{UadL} it is  obvious that
the renormalization of the g-factor is also reflected in the
renormalization of the relevant spin-orbit length scales,
$l_\so\to \tilde{l}_\so\equiv\sqrt{|\det\uu{\tilde{\lambda}}_\so|}$.
This renormalized $\tilde{l}_\so$ has been shown in  Fig. \ref{fig:dressed} as a function of
the ratio $x_0/l_\so$. In small quantum dots, $x_0/l_\so\ll
1$, one has $z_{\mu\nu}=\delta_{\mu\nu}$, and therefore
$\uu{\tilde\lambda}_\so^{-1}=\uu\lambda_\so^{-1}$, in
agreement with the assumptions of Sec.~\ref{sec:concepts}.

\subsubsection{Connection with perturbation theory}

The preceding  discussion  relied on the  specific model
of a parabolic quantum dot, which enabled us to construct
the instantaneous eigenstates of the Hamiltonian
for arbitrarily large but slowly changing displacements,
$\vc{R}_\mathcal{C}\sim \vc E$. An alternative approach to construct
$U_\mathrm{ad}$ that is valid for an arbitrary shape of the
confining potential is to use perturbation theory in the
driving term $H_\mathrm{int}=e\vc E(t)\vco r$ of Eq.\
(\ref{H}) to determine the instantaneous eigenstates of
$H(t)$\cite{sanjose06}.  This perturbative approach has the disadvantage of
breaking down for displacements $\vc{R}_\mathcal{C}$
comparable to the typical dot size $x_0$. However, in a
conventional setup of single quantum dots defined by
lithographic gates,  one is usually constrained to small
displacements $|\vc{R}_\mathcal{C}|\ll x_0$ such  that the
perturbative approach remains useful.

In the absence of external magnetic fields the perturbation
$H_\mathrm{int}=e\vc E(t)\vco r$ does
not connect states within the same doublet due to
time reversal symmetry.  One can then
approximate the lowest energy instantaneous eigenstate $|\tau_0(t)\rangle$
 as
\begin{eqnarray}
|\tau_0(t)\rangle&\equiv& \mathcal{W}_{\rm pert}(t)|\tau_0\rangle\;,
\nn
\\
 \mathcal{W}_{\rm pert}(t)
&\approx & \sum_\tau \Bigl(
{|\tau_0\rangle} \langle \tau_0|
+ eE_\mu(t)\sum_{n\ne 0}
\frac{|\tau_n\rangle \langle \tau_n| \hat{r}_\mu}
{E_0-E_n}\Bigr)\;,
\end{eqnarray}
where for brevity we have denoted by $E_n$ the energy of doublet
$n$ at time $t=0$, and $|\tau_n\rangle\equiv|\tau_n(0)\rangle$.
Then the effective adiabatic Hamiltonian, describing the
evolution of the center of motion wave function
$|\tilde \Psi(t)\rangle_{\rm pert}
\equiv \mathcal{W}_{\rm pert}^+(t)| \Psi(t)\rangle $
within the ground state multiplet, reads
\be
\tilde H_\mathrm{eff} \approx
\frac{i}{2}e^2\bigl (\dot
{\vc E}\times\vc
E\bigr)_z \sum_{n>0} P_0
\frac{\hat{x}P_n\hat{y}-\hat{y}P_n\hat{x}}{(E_n-E_0)^2}P_0 \; ,\nn
\ee
where the operators  $P_n$ project to the $n$'th doublet state
of the unperturbed Hamiltonian ${\mathcal H}(0)$.
Furthermore, in this equation we dropped full differential terms involving
$\dot E_\mu E_\nu+E_\mu\dot E_\nu$ that give no contribution  for
closed paths, and we denoted by $\hat{x},\hat{y}$ the
components of the electron position operator.
The above expression coincides with the $\vc B=0$ limit of the results
obtained previously using  a more general perturbative approach in
Ref.~\onlinecite{sanjose06}.\footnote{We denoted this term  by $\vc
C^{(3)}_{\mu\nu}$ in Ref.~\onlinecite{sanjose06}} Treating the
general case, $\vc B\neq 0$, within this approach
 requires some care since the appropriate perturbation theory is
quasi-degenerate~\cite{sanjose06,sanjose07}.

While it is not obvious, the above perturbation expansion
approach and the displacement operator results presented in the previous
subsection are equivalent.
Due to the slightly different choice of the transformation
$\mathcal{W}^+(t)$ the two approaches yield different adiabatic evolution operators,
however, both Hamiltonians
 describe  the same physical pseudospin precession, and physical observables are independent of this choice of basis.
To illustrate this, let us compare   the two
matrices $\tilde U_\mathrm{ad}$ after a path that starts and ends
at the origin. For a closed path $\vc R_\mathcal{C}(t)$,
the perturbative approach yields a pseudospin
transformation that can be expressed as
\begin{eqnarray}
\tilde U^\mathrm{pert}_\mathrm{ad}&=& {\rm P}
\exp{\left[-\frac{i}{2 \;x_0^2} \int_{\mathcal C}(d\vc{R}_\mathcal{C}\times\vc{R}_\mathcal{C})_z \; \vc{C}^\mathrm{pert}_{xy}\cdot\vco\tau\right]}
\nn\\
\vc{C}^\mathrm{pert}_{xy}\cdot\hat{\tau}_{\tau\tau'}&=&i\frac{\omega_0^2}{x_0^2}\sum_{n>0}\frac{\langle\tau_0|\hat{x}P_n\hat{y}-\hat{y}P_n\hat{x}|\tau'_0\rangle}
{(E_n-E_0)^2}\label{Upert}
\end{eqnarray}

As we show in Appendix~\ref{ap:Dressing},
 for the  particular case of a parabolic dot
sum rules imply that
$\vc{C}^\mathrm{pert}_{xy}$ simplifies to
\begin{eqnarray}
\vc{C}^\mathrm{pert}_{xy}\cdot\hat{\sigma}_{\tau\tau'}&=&-ix_0^2 \langle\tau_0|\hat{p}_xP_0\hat{p}_y-\hat{p}_yP_0\hat{p}_x|\tau'_0\rangle
\label{Cpertpar}
\end{eqnarray}

On the other hand, the exact displacement operator approach of the previous
subsection   yields
\begin{eqnarray}
\tilde U^\mathrm{par}_\mathrm{ad}&=&{\rm P}
\exp{\left[i\int_\mathcal{C}d\vc{R}_\mathcal{C}\cdot P_0\vco p P_0\right]}\;.
\end{eqnarray}
For a closed path with a small enclosed area we can
approximate this expression by expanding to second order
in the exponent and re-exponentiating as
\begin{eqnarray}
\tilde U^\mathrm{par}_\mathrm{ad}&=&
 {\rm P}\; \exp\Big[i\int_\mathcal{C}(d\vc{R}_\mathcal{C}\times \vc
R_\mathcal{C})_z \times
\\
& & \times \frac{i}{2}\left(P_0\hat{p}_xP_0\hat{p}_y
P_0-P_0\hat{p}_yP_0\hat{p}_x P_0\right)\Big]
\label{Upar} \, .
\end{eqnarray}
Eqs. (\ref{Upert}), (\ref{Upar}) and
Eq. (\ref{Cpertpar}) imply that both approaches yield
identical results. Furthermore, in the case of a parabolic dot
we can show using the alternative form for
$\tilde U_\mathrm{ad}$, Eq.~(\ref{UadL}), that for small closed paths
the evolution operator $U_\mathrm{ad}$ of the
unshifted states, $|\Psi(t)\rangle$ is approximately given as
\begin{equation}
 U_\mathrm{ad} \approx  \tilde U_\mathrm{ad}=\exp\left(-\frac{i}{2}\frac{4A}{\tilde{l}_\so^2}\hat{\sigma}_z\right)
\, ,\label{UadA}
\end{equation}
where $A=\frac{1}{2}\int_C (d\vc R_\mathcal{C}\times\vc
R_\mathcal{C})_z$ is displacement area spanned by the path.
This generalizes the semiclassical result of the previous section
to the case of arbitrary spin-obit coupling.

\subsection{Treating the electromagnetic field quantum-mechanically \label{sec:QeQf}}

In the first part of this section
we assumed that the time-dependent electric field
acting on the electron is classical.
However, if the typical frequency of the field reaches the
temperature then the electric field must be treated quantum-mechanically.
To treat the quantum fluctuations of the electromagnetic field,
we shall employ a hybrid formalism, where we
describe the fluctuations of the electromagnetic field
within the path integral approach, while
 the quantum dot shall be treated  within an operator formalism.
Here we only summarize the results, the
details of the calculation are presented in Appendix~\ref{ap:quantumfields}.

For simplicity, we consider an electron spin in a parabolic quantum
dot and dipolar electric fields as in subsection~\ref{subsub:parabolic}.
Then the adiabatic evolution operator that describes the evolution
of the dot state $|\Psi(t)\rangle$ for given initial $\vc E_i$ and final
$\vc E_f$ states of the electromagnetic environment within the ground doublet
is given by
\begin{eqnarray}
&&\langle \vc E_f| U_\mathrm{ad}(t)|\vc
E_i\rangle= \label{Uadquantum}
\\
&&\phantom{n}\mathcal{W}_{\vc E_f}  \Bigl\{\int_{\vc E_i}^{\vc E_f}\mathcal{D}[\vc
E]\, e^{-iS'_B}
{\rm T}\;e^{-i\int_0^tdt'\left[H_Z+H_G(\dot{\vc
      E}(t'))\right]}\Bigr\}\mathcal{W}_{\vc E_i}^+,
\nn\\
&&
\phantom{nnn}H_Z=\frac{g}{2}\mu_B\vc B\, P_0 \vco\sigma P_0\,
\label{eq:H_Z}
\\
&&\phantom{nnn}H_G(t)=\dot{\vc R}_\mathcal{C}\;
\uu{\lambda}^{-1}_\so    P_0    \vco \sigma P_0 =
- \frac{e\dot {\vc E}}{m\omega_0^2} \uu{\lambda}^{-1}_\so    P_0    \vco \sigma P_0
\;.
\label{eq:H_G}
\end{eqnarray}
Here the functional integral is performed over all possible
fluctuations of the bath,  $\vc E(t)$,   each of them corresponding
to a different path $\vc R_\mathcal{C}(t)$ between times $0$ and $t$.
We also allowed for the presence of an in-plane magnetic field $\vc B$, small
compared to the level spacing of the dot, $\sim \omega_0$.
The adiabatic displacement operators $\mathcal{W}=e^{i\vco pe\vc
E/m\omega_0^2}=e^{-i\vco p\vc R_\mathcal{C}}$  in Eq.~\eqref{Uadquantum}
transform the wave function
of the dot to the center of motion basis,  $|\Psi(t)\rangle\to  |\tilde
\Psi(t)\rangle$, where the evolution is described by
Eq.~\eqref{U}, projected by the projector $P_0$ to the lowest
lying two eigenstates
of  ${\mathcal H}(0)$, in the absence of the magnetic field.

Thus the pseudospin evolution under
a quantum electric field is the coherent sum of
classical evolutions over all  possible field
fluctuations weighted by the action of the decoupled bath,
$S_B'$. We remark here that $S_B'$
is not the non-interacting bath action, $S_B$, but
contains a
correction $\Delta H_B=-e^2|\vc
E(t)|^2/(2m\omega_0^2)$ due to the back reaction of the dot
(see Appendix~\ref{ap:quantumfields}). For a Gaussian $S'_B$
we thus mapped the evolution of the spin to the
well-studied spin-boson model, where, however, external bosonic fluctuations
are coupled both  to the x and y components of the spin.
\cite{CastroNetoPRL03,NovaisPRB05}

To describe dephasing and spin relaxation at the fully
quantum-mechanical level we generalized the formulas above
 to the evolution
of the reduced density matrix for the pseudospin in the center of motion
 basis,
\begin{equation}
\tilde \rho_D(t)_{\tau\tau'}\equiv \langle\tau_0|\Tr_B\left[
\hat{\mathcal{W}}^+\rho(t)\hat{\mathcal{W}}\right]|\tau'_0\rangle\label{rhodef}
\end{equation}
where ${\rm Tr}_B$ stands for a trace over the bath degrees of
freedom,  $\rho(t)=U(t)\rho(0)U^+(t)$ is the density
matrix of the complete dot-bath system,
and $\hat {\mathcal{W}}\equiv \exp(-i\vco p\cdot\vco R_\mathcal{C})$.
Note that now $\vco
R_\mathcal{C}=-e\vco E/m\omega_0^2$ is an operator instead
of a c-number, and that  $\hat {\mathcal{W}}
=\exp(-i\vco p\cdot\vco R_\mathcal{C})$
acts both on the bath and on the quantum dot.
Then spin relaxation and decay corresponds to the diagonal and
off-diagonal  components of $\tilde \rho_D(t)_{\tau\tau'}$,
respectively. The initial state $\rho(0)$ of the system should not affect
the dynamics at long times \cite{Hakim85},
 so we will
assume the dot-plus-bath system to start at $t=0$ in a well
defined ground doublet and the bath in thermal equilibrium,
$\tilde \rho(0)\equiv \hat{\mathcal{W}}^+\rho(0)\hat{\mathcal{W}}=\tilde
\rho_D(0)\otimes\rho_B(0)$.
This choice is technically convenient and physically
describes a definite initial state  with slightly entangled
bath and dot states.
Then the evolution of $\tilde \rho_D(t)$ can be described as
\begin{equation}
\tilde \rho_D(t) = \left \langle{\rm T}_K
\left[e^{-i\int_K dz \left(H_Z+H_G\right)}\tilde \rho_D(0)\right]\right\rangle_B\;,
\label{rhot}
\end{equation}
where now $z$ runs along the  usual double branch time
contour $K$ running from $z_+=0$ to $t$ and then back to
$0$, as depicted in Fig.~\ref{fig:diagrams} in Appendix~\ref{ap:diagramatic},
 and $T_K$
is time ordering along this contour. Here
$\langle \dots \rangle_B$ denotes the average over the electromagnetic field
fluctuations along the Keldysh contour,
$$
\int \mathcal{D}[\vc E]
\;\langle \vc E_+{(0)}|\rho_B(0)|\vc E_-{(0)}\rangle\;
e^{-iS_B'[\vc E_+]+iS_B'[\vc E_-]}\; \dots,
$$
where $\vc E_+$ and $\vc E_-$ denote the electric field along the
upper and lower branches of the contour and satisfy the boundary
condition, $E_+(t)=E_-(t)$.

Unfortunately, one cannot integrate out the electric
field exactly.  However,  one can use the systematic
diagrammatic approach of Ref.~[\onlinecite{Schoeller94}]
to do  perturbation theory in $H_G(t)$
and  obtain the relaxation rates within a Markovian approximation.
\cite{Makhlin02} The results of this calculation are presented in the
following Section.

\section{Results for relaxation and dephasing\label{sec:results}}

We are now ready to summarize the results that are obtained by the formalism
developed in the previous section. Some technical details of the
calculation are presented in  Appendix \ref{ap:diagramatic}.
We remark here the present systematic approach confirms what had been
derived within the perturbative formalism developed in
Ref.~\onlinecite{sanjose06}. Unlike in said work, where numerical results for $T_1$ were computed in a limited set of cases, the results for $T_1$ and $T_2$ presented in this section are fully analytical, which therefore allows for an explicit analysis of the decay rate dependence with all relevant physical parameters, such as magnetic field orientation, spin-orbit coupling or quantum dot size.

We have computed the spin relaxation ($T_1^{-1}$) and dephasing ($T_2^{-1}$) rates of confined electrons subject to an in-plane (renormalized) magnetic field $\vc B$ at
an angle $\theta$ with respect to direction $[100]$.
To  fourth order in the spin-orbit coupling we obtain
\begin{eqnarray}
\frac{1}{T_1}&=&2(m x_0)^2(\alpha^2+\beta^2-2\alpha\beta\sin 2\theta)\omega_B^2\coth\frac{\omega_B}{2k_BT}A(\omega_B)\nn\\
&+&2(mx_0)^4\left[\left(\alpha^2+\beta^2\right)^2+4\alpha^2\beta^2\cos
4\theta\right]F^+(\omega_B)\nn\\
&+&2(mx_0)^4  (\alpha^2 -\beta^2)^2
F^-(\omega_B)\label{T1f}
\;,
\\
\frac{1}{T_2}&=&\frac{1}{2T_1}+(m x_0)^4(\alpha^2+\beta^2-2\alpha\beta\sin
2\theta)^2F(\omega_B)
\;.
\label{T2f}
\end{eqnarray}
Here $\omega_B=-g\mu_B|\vc B|$, $x_0$ is the quantum dot size (assumed parabolic)
and the functions $F^\pm $ and $F$
are defined as
\begin{eqnarray}
F^+(\omega_B)&=&\omega_B^2\int_{-\infty}^\infty
\frac{d\tilde\omega}{8\pi}
\frac{A\left(\frac{\omega_B+\tilde\omega}{2}\right)A\left(\frac{\omega_B-\tilde\omega}{2}\right)}{1-\frac{\cosh(\tilde\omega/2K_BT)}{\cosh(\omega_B/2K_BT)}},\nn\\
F^-(\omega_B)&=&\int_{-\infty}^\infty
\frac{d\tilde\omega}{8\pi}\tilde\omega^2
\frac{A\left(\frac{\omega_B+\tilde\omega}{2}\right)A\left(\frac{\omega_B-\tilde\omega}{2}\right)}{1-\frac{\cosh(\tilde\omega/2K_BT)}{\cosh(\omega_B/2K_BT)}},\nn\\
F(\omega_B)&=&\int_{-\infty}^\infty
\frac{d\tilde\omega}{8\pi}\tilde\omega^4A(\tilde\omega)^2\mathrm{csch}^2\left(\frac{\tilde\omega}{2K_BT}\right)\nn\\
&&\times\mathrm{Re}\left[\left(\frac{1}{\omega_B-\tilde\omega-i
      0^+}+\frac{1}{\omega_B+\tilde\omega+i 0^+}\right)^2\right],\nn
\end{eqnarray}
with the spectral function $A(\omega)$
of the dimensionless electric field, $ex_0\vc E/\omega_0$,
defined in Appendix~\ref{ap:diagramatic}. Functions $F^{\pm}$ relate to the $F^K_{1,2}$ used in Appendix~\ref{ap:diagramatic} by $F^{\pm}=(F^{K}_1\pm F^{K}_2)/2$.
Note that both $F^+(\omega_B)$ and $F(\omega_B)$ vanish
for $\omega_B\rightarrow 0$, while $F^-(0)$ remains
finite. Therefore, this latter part of the fourth order contribution can be
identified as the Berry phase contribution. Indeed, this expression is
identical to the one we obtained previously in Ref.~[\onlinecite{sanjose06}].
Note also that this term is
isotropic and does not depend on the direction of the magnetic field, while
all other second and fourth order contributions do so for
$\alpha,\beta\ne 0$.
In the formulas above
we assumed that the spectral functions of the components of the
 dimensionless electric field $ex_0\vc
E/\omega_0$ are isotropic,   $A_x(\omega) =A_y(\omega) = A(\omega)$.

The spectral function $A(\omega)$ gives the density of
electromagnetic excitations that contribute to dephasing, and it depends on the specific source
of electromagnetic fluctuations.
Important sources of
electric field fluctuations are piezoelectric phonons
\cite{Khaetskii00} and at low magnetic fields
 Ohmic charge fluctuations\cite{sanjose06},
$$
A(\omega)=A_\mathrm{ph}(\omega)+A_\Omega(\omega).
$$

 For the case of
piezoelectric phonons an estimate of the spectral function
$A_\mathrm{ph}(\omega)$ of the induced dimensionless
electric field is outlined in Appendix \ref{ap:phonons}.
For GaAs/GaAlAs heterostructures we obtain
$A_\mathrm{ph}(\omega)=\omega^3\lambda_\mathrm{ph}x_0^2/\omega_0^2$,
with $\lambda_{\mathrm{ph}}=2.5\cdot
10^{-5}\mathrm{K^{-2}nm^{-2}}$. At higher magnetic fields (frequencies)
these fluctuations give the dominant contribution to spin
relaxation/dephasing.

At low magnetic fields (low
frequencies), on the other hand,  Ohmic charge fluctuations of the electric
environment of the 2DEG near the dot
are expected to dominate. For
these, the spectral function is
$A_\Omega(\omega)=\lambda_\Omega\omega/\omega_0^2$ for
Fermi liquid leads, with $\lambda_\Omega\sim
(e^2/h)\mathrm{Re}[Z]$, $Z$ being the impedance of the leads.
It is rather difficult to compute the
 exact  value of  $\lambda_\Omega$, since it depends on the precise geometry
 of the leads and one must also
take into account how the equilibrium electric field
fluctuations inside the 2DEG
extend to the quantum dot, and to what extent those
fluctuations can be screened. It is not unreasonable,
however, for typical sheet resistances of
$10^2-10^3\Omega/\Box$ to assume $\lambda_\Omega\sim
10^{-3} - 10^{-4}$.

In our calculations we have neglected the feedback
correction $-e^2|\vc E|^2/(2m\omega_0^2)$ that changes the
action $S_B\to S'_B$. In the case
of the phonon bath, the effect of this term is to make
phonons somewhat softer close to the dot.
However, this polaron-type effect  should
be small, since the mass of the electron is negligible
compared to the atomic masses.
The feedback correction might be
more important  for the Ohmic bath, but will only
affect $\lambda_\Omega$, not the Ohmic character of the
bath.

\begin{figure}
\includegraphics[width=6.5 cm]{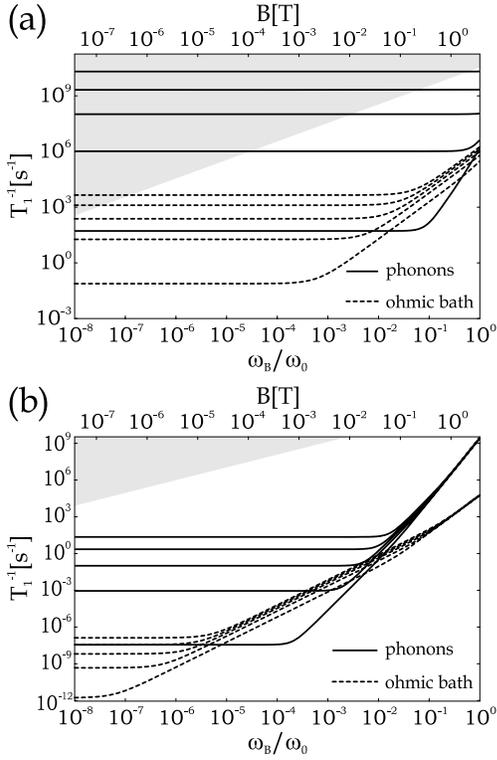}
\caption{Relaxation rates due to piezoelectric phonons and
Ohmic fluctuations in two scenarios, both as a function of
the magnetic field $\vc B$ applied along the [100]
direction. For the strengths of Rashba and Dresselhaus
spin-orbit coupling a ratio $\alpha/\beta=4$ is assumed \cite{Averkiev06}.
In (a) we plot the unfavorable scenario for quantum
information precessing, with $\lambda_\Omega=5\cdot
10^{-3}$, $l_\so=1500\mathrm{nm}$ and a large dot $x_0=115
\mathrm{nm}$ ($\omega_0\approx 1 \mathrm{K}$). In (b) we
plot the rates for favorable conditions
$\lambda_\Omega=10^{-4}$, $l_\so=3000\mathrm{nm}$ and
$x_0=36\mathrm{nm}$ ($\omega_0=10 \mathrm{K}$). Lines range
from a bath temperature of $100 \mathrm{mK}$ (lower rates)
to $900 \mathrm{mK}$ (higher rates) in steps of
$200\mathrm{mK}$. Shaded in gray is the region
$T_1^{-1}>\omega_B=-g\mu_B|\vc B|$ for which the rotating wave approximation fails and the Zeno regime sets in (see Sec. \ref{sec:clafluc}). Note
the extreme dependence of relaxation rates on the specific
conditions. \label{fig:rates}}
\end{figure}

\begin{figure}
\includegraphics[width=7 cm]{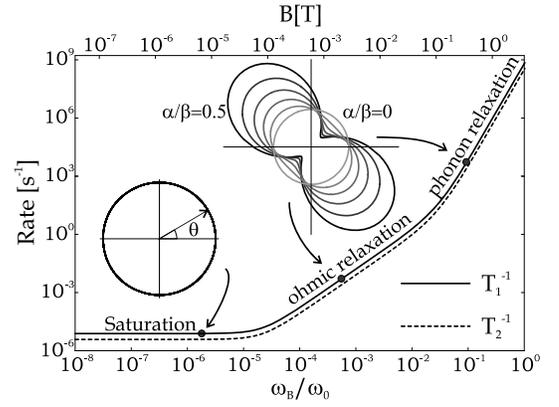}
\caption{Total relaxation and decoherence rates for a
$51\mathrm{nm}$ dot ($\omega_0=5\mathrm{K}$) at
$T=100\mathrm{mK}$. Other parameters are
$l_\so=3\mathrm{\mu m}$, $\alpha/\beta=4$, $\theta=0$, $\lambda_\Omega=10^{-3}$. Three regimes are clearly
visible. In the inset we give a polar plot of the dependence of the rates with magnetic field angle $\theta$. This angular dependence is negligible in the saturated regime, and of
the form
$T^{-1}\propto\alpha^2+\beta^2-2\alpha\beta\sin2\theta$ at
higher fields.
\label{fig:angular}}
\end{figure}

\begin{figure}
\includegraphics[width=7 cm]{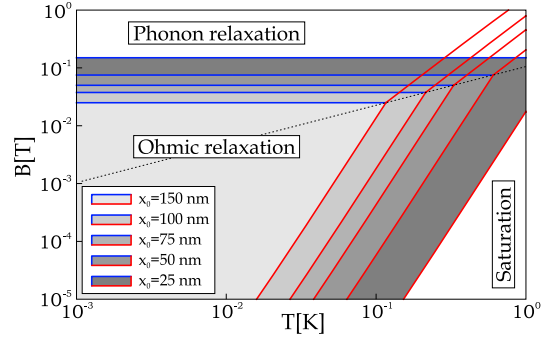}
\caption{(Color online) Phase diagram of the three
relaxation regimes at $\lambda_\Omega=10^{-3}$,
$\alpha/\beta=4$, $l_\so=3\mathrm{\mu m}$ and $\theta=0$.
A window of ohmic fluctuations - dominated relaxation and
dephasing opens up at lower
temperatures. The saturation regime in this plot is dominated by phonon fluctuations, although regions exist (not shown) at lower fields and temperatures where the saturation is mainly due to \emph{Ohmic} fluctuations.
\label{fig:phasediagram}}
\end{figure}

Implicit in the derivation of Eqs. (\ref{T1f}) and (\ref{T2f}) is the assumption that the effect of the thermal bath leads to an adiabatic evolution of the spin in the quantum dot. In physical terms one can anticipate that this implies a condition for the bath temperature $k_B T\ll \omega_0$, so that the dot is not heated above the ground doublet by the bath fluctuations. More precisely, the classical adiabatic condition $\dot{\vc R_\mathcal{C}}/x_0\ll \omega_0$ given in Sec. \ref{sec:geomphases}, translates in this context into the relation $(k_BT)^3 A(k_BT)\ll \omega_0^2$. For typical parameters in the case of the ohmic bath this means indeed $0.2 k_B T\ll \omega_0$. For the piezoelectric phonon bath this implies $6\left(T[\mathrm{K}]\right)^{1/5} k_B T\ll \omega_0$, where $T[\mathrm{K}]$ is the bath temperature in Kelvin. Non-adiabatic corrections are beyond the scope of this work, although they could in principle be taken into account in the calculation of Appendix \ref{ap:diagramatic}.

The first (second order) term in Eq.~(\ref{T1f}) is proportional
to  $\omega_B A(\omega_B)$ at small Larmor frequencies.
Concentrating on this term one concludes that the relaxation and dephasing
rates vanish for  $\omega_B\to 0$~\cite{Khaetskii00,Stano06,Semenov06}.
The second and third terms  in  Eq. (\ref{T2f}), which are
of fourth order in the spin-orbit couplings,
involve a convolution of the spectral function at
frequencies up to the temperature scale, and
$F^-(\omega_B)$  does not
vanish at low fields and non-zero temperature.
Therefore the relaxation rates saturate
both for the phonon and the Ohmic baths
as the field is lowered.
We remark that the
 sum of the terms $F^+$ and $F^-$  resemble the
 formula, Eq.~\eqref{eq:4order_estimate},
anticipated  in Sec. \ref{sec:clafluc}, although the full quantum
treatment given here is necessary to
identify precisely the functions
$C_{\perp}$  and  $C_{\parallel}$ in  Eq.~\eqref{eq:4order_estimate} and to
arrive at consistent quantitative predictions.

In Fig. \ref{fig:rates}, we plot the relaxation rates
induced by  the Ohmic and phonon baths separately
for two different sets of parameters. For small
magnetic fields $\hbar\omega_B\ll k_BT$ (but still above the Zeno regime) the relaxation rates reach a clear saturation regime, with values
\begin{eqnarray}
&&T_1^{-1}\approx 2T^{-1}_2\approx 4\frac{(k_BT)^5}{\omega_0^4}\left(\frac{x_0}{l_\so}\right)^4\label{satval}\\
&&\times\left[C_4 \lambda_\Omega^2+2C_6
(k_BTx_0)^2\lambda_\Omega\lambda_\mathrm{ph}+C_8(k_BTx_0)^4\lambda_\mathrm{ph}^2\right]\nn
\end{eqnarray}
where the  numerical constants equal
equal $C_4=8\pi^3/15\approx 16.5$, $C_6=32\pi^5/21\approx 466$
and $C_8=128\pi^7/15\approx 2.58\cdot 10^4$.

As the field is increased, the second order terms begin to
dominate. At low enough temperatures Ohmic fluctuations  become
dominant first, since their weight grows faster at low
fields than the phonon contribution.
Thus   below a certain
temperature there is a  window of magnetic fields in
which the relaxation rates have a scaling form characteristic of
an Ohmic bath,
$$
T_1^{-1}\approx 2T_2^{-1}\propto
\omega_B^3 \;\coth\frac{\omega_B}{2k_BT}
\, .
$$
At still larger values of the magnetic field the phonon fluctuations
dominate the second order relaxation channel,
and the rates assume a 'phononic' scaling form,
$$
T_1^{-1}\approx 2T_2^{-1}\propto\omega_B^5 \; \coth\frac{\omega_B}{2k_BT}
\,
.
$$
 These cross-overs are shown in Fig. \ref{fig:angular}
for both relaxation and dephasing rates.
The different
regimes are  summarized in a 'phase diagram', which we show
for typical parameters in Fig.~\ref{fig:phasediagram}.

As noticed in Ref. \onlinecite{Golovach04},
$T_1^{-1}=2T_2^{-1}$ up to  second order in the
coupling $x_0/l_\so$ (i.e. at high fields).
This relation is violated by the fourth order terms due to the $F$ contribution in (\ref{T2f}),
but is again restored in the saturation regime,
where the fourth order term $\sim F$ not considered in Ref.~\onlinecite{Golovach04}
vanishes.

For $\alpha,\beta\ne 0$  second order terms that
 dominate  the high field behavior  have a
strong dependence on $\theta$:
$T^{-1}\propto\alpha^2+\beta^2-2\alpha\beta\sin(2\theta)$.
Their contribution to
the relaxation rate is enhanced for fields along [110],
especially as $\beta$ approaches $\alpha$ for a fixed
$l_\so$ (see Fig.~\ref{fig:angular}).
This is easy to understand by
looking back at Fig.~\ref{fig:rolling}(b). The
relaxation of the spin occurs due to dot displacements
along directions that flip the spin, which for spins along
the [110] direction ($\theta=3\pi/4$) are displacements
along [110] itself (recall the geometric interpretation as
a rolling sphere). As we see from Fig.~\ref{fig:rolling}(b) such angles are the most effective
to induce spin flips, so the  relaxation and also
decoherence rates increase for fields in those directions,
especially for the highly anisotropic case
$\alpha\sim\beta$. In contrast, as is obvious from
Eq.~(\ref{satval}) and Fig.~\ref{fig:rolling}(b), at low fields the rates are dominated by
the geometric term that is  independent of $\theta$.

\section{Electric  spin manipulation\label{sec:manipulation}}

In the previous sections we investigated how
in the presence of spin-orbit interaction
stochastic and fluctuating fields lead to decay and
decoherence of electron spin states.
In the present section we discuss how one can use this
effect in a constructive way to
{\em control }  spins purely by electric fields.
For this purpose one should
displace the quantum dot which confines the electron.
Unfortunately, under realistic conditions such displacements
are rather small compared to $l_{so}$.
Nevertheless, a series of small
closed paths  can be designed to take the spin of the
confined electron to an arbitrary final state.
In this way one can realize an all-electrical universal single qubit gate, under the realistic condition that the scattering mean free path in the 2DEG (typically in the $\mu \mathrm{m}$ range) is much larger than the confinement lengthscale $x_0$.
Another option that we shall discuss in this section is
to move an electron in a system of quantum dots
controlled by gate voltages, and in this way manipulate its spin state.

Closed trajectories of the electron
(in a confining dot) covering an area $A$ induce a spin precession
approximately given by Eq. (\ref{UadA}) around the $\hat{z}$ axis,
in the same way as if a
constant magnetic field was applied in this direction.
In order to induce arbitrary spin rotations, e.g. spin flips,
we have to rely on more complicated paths. An example is
a path composed of  sum of closed loops of period $T_f$ and area $A_f$
and another, much slower, closed trajectory of period $T_s$
and area $A_s$, i.e., a spirograph-type of path. One can
demonstrate that by properly choosing the relation between
frequencies and trajectories,  after a long enough driving
time the spin can be driven to an arbitrary final state.
The optimal relation for a spin-flip operation
is $T_f/T_s=2A_f/(\pi\tilde{l}_\so^2)$. Due to the
adiabaticity requirement, however, the minimal
spin-flip time using this method for realistic values of
the maximum displacement becomes several orders of
magnitude slower than current flip times achieved using ESR
techniques \cite{Koppens06}. The operation time can
be reduced by using heterostructures with larger spin-orbit
couplings, such as InAs.
It can also be reduced substantially if we substitute the effect of the
fast path component by an equivalent external static
magnetic field along the $\hat{z}$ direction, in which case
the technique resembles closely previously proposed ac
electric-field generalizations of ESR techniques
\cite{Duckheim06}.

One can, however,  also  use different methods to transport a
confined electron over  distances comparable or greater than the spin-orbit
length $l_\so\sim 3\mathrm{\mu m}$.
Surface acoustic waves, e.g., have been used to move electrons over large
distances and to rotate their spins \cite{Stotz04}.
Another possibility involves a ring of
several tunnel-coupled quantum dots spanning distances
comparable to $l_\so$.
 The electron  can then be adiabatically shifted
around the ring  by  appropriate  time-dependent gate voltages.
As we show below, such a manipulation can result in a large spin rotation,
and thus provide a
completely different principle than ESR, since no resonant
AC fields would be involved. For this reason it could be
expected to enable relatively fast coherent spin
manipulation, since, unlike ESR, a single pumping cycle
could be enough to induce a complete spin flip.

The discussion of pumping around the ring of dots can be
reduced to sequential pumping processes between adjacent dots.
We shall therefore
analyze the  precession of the (pseudo)spin  of an electron
that is transferred  between two dots  of sizes and separation
smaller  than $l_\so$. We further assume that
a strong barrier between the dots remains present
at all times during the pumping process, and
use a tight-binding approximation. In this spirit, we
write the  Hamiltonian of the
double dot structure as
\begin{equation}
H(t)=\frac{{\vco p}^2}{2m}+\sum_{\alpha = L,R} V_\alpha(\vco r)
+\frac{1}{m}\vco
p\uu\lambda_\so^{-1}\vco{\sigma} + V_{\rm ext}(\vco r ,t)\;,
\end{equation}
where $V_\alpha(\vco r) \equiv V(\vco r-\vc r_0^\alpha)$ denotes the confining
potential of the two dots at positions ${\vco r}_0^L$ and ${\vco r}_0^R$, and
the potential $V_{\rm ext}$ is generated by the external gate
voltages, assumed to be small. Within the tight binding
approximation  the low energy states of the double dot
system  are a linear combination  of the ground state wavefunctions
$\{|\psi^L_{0\pm}\rangle,|\psi^R_{0\pm}\rangle\}$
 of the isolated dots described by the Hamiltonians
\begin{equation}
H_{L/R} \equiv\frac{\vco{\hat{p}}^2}{2m}+ V_{L/R}+\frac{1}{m}\vco
p\uu\lambda_\so^{-1}\vco{\sigma}
\end{equation}
In the considered limit
 the SO coupling can be treated perturbatively, and  the Hamiltonians
$H_{L/R}$  can be partially diagonalized
by the unitary transformation
$\mathcal{Z}_{L/R}=e^{i {M}_{L/R}}$, with
\begin{eqnarray}
\mathcal{M}_\alpha&=&-(\vco
r-\vc r^\alpha_0)\uu\lambda_\so^{-1}\vco{\sigma}
\nn\\
 \mathcal{Z}_\alpha^+ H_\alpha
\mathcal{Z}_\alpha
&=&\frac{p^2}{2m}+V(\vc r-\vc r^\alpha_0)-\frac{1}{m
l_\so^2}L_zS_z+\mathcal{O}^3(\uu\lambda_\so^{-1}r).\nn
\end{eqnarray}
Here $l_\so=\sqrt{|\det\uu\lambda_\so|}$ and $L_z$ is the
angular momentum with respect to the center of the dot.
Let us now assume that the confining potentials $V_\alpha$
are cylindrically symmetrical and also that the ground states
of $H_\alpha$ are in the $L_z=0$ sector. Then the ground state of the individual
dots can be approximated, to order $x_0^2/l_\so^2$, as
\begin{equation}
|\phi^\alpha_{0\sigma}\rangle\approx
e^{iM_\alpha}|\Psi^\alpha_{0} \rangle\otimes|\sigma\rangle
\label{eq:appr_wave_function}
\end{equation}
in terms of the spinor $|\sigma\rangle$ and the orbital
eigenstate $|\Psi^\alpha_{0} \rangle$ of $H_\alpha$ with
the spin-orbit coupling set to zero.  This allows us to evaluate the
$4\times 4$ matrix for the truncated $H(t)$ and express it as
\begin{equation}
H(t)=\left(\begin{array}{cc}H_{LL}&H_{LR}\\H_{RL}&H_{RR}\end{array}\right)\;,
\end{equation}
with the submatrices given by $(H_{\alpha\beta })_{\sigma\sigma'} =
\langle\phi^{\alpha}_{0\sigma}|H(t)|\phi^\beta_{0\sigma'}\rangle$.
Apart from a trivial overall shift of the energy, the diagonal blocks can be
written as
\begin{equation}
H_{LL}= -H_{RR}=\frac{v(t)}{2}
\left(\begin{array}{cc}1&0\\0&1\end{array}\right)\;,
\end{equation}
 $v(t)$ being the potential
difference between the dots. However,
spin-orbit coupling generates a spin texture for the  dot eigenstates,
and results in a non-trivial spin-mixing in the
in the hopping submatrix, $\alpha\neq\alpha'$:
\begin{eqnarray}
H_{LR}&=&H_{RL}^+=\langle\phi^{L}_{0\sigma'}|H(t)|\phi^R_{0\sigma}\rangle\label{Hpump}\\
&\approx&\langle\Psi^{L}_{0}|H(t)|\Psi^R_0\rangle\langle\sigma'|e^{
-i \;\delta \vco r\cdot\uu\lambda_\so^{-1} \vco \sigma}
|\sigma\rangle\;.
\end{eqnarray}
where $\delta\vc r=\vc r_R-\vc
r_L$ is the vector connecting the two dots.
In the second line we used the expression \eqref{eq:appr_wave_function}
and exploited the fact that the integrals pick up their major contributions from
the regions $\vc r\approx \vc r_L$ and $\vc r\approx \vc r_R$.
We thus obtain,
\begin{eqnarray}
H_{LR} \approx \frac{\Delta(t)}{2}\left(\begin{array}{cc}
\cos(\rho
)&
ie^{-i\phi}\sin(
\rho
)\\
ie^{i\phi}\sin(
\rho
)&
\cos(
\rho
)
\end{array}\right)\nn
\end{eqnarray}
where $\rho= |\uu\lambda_\so^{-1}\cdot\delta \vc r|$ is essentially the tunneling distance in
units of the spin-orbit length and
$\Delta (t)$ the spin-independent hopping
integral $\langle\Psi_0^L|\dots|\Psi_0^R\rangle$ above. The angle
$\phi$ characterizes the hopping direction and is defined through the relation
$\delta\vc r\cdot\uu\lambda_\so^{-1}=\rho (\cos\phi, \sin\phi)$.
The SU(2) operator
$e^{-i\delta\vc
r\cdot\uu\lambda_\so^{-1}\cdot\vc\sigma}\propto H_{RL}$ is
the same we obtained in Eq. (\ref{UadL}) for the geometric
spin precession along a straight path connecting the two
dots.

Diagonalizing this Hamiltonian matrix we obtain  the
instantaneous eigenstates
$\{|\phi_{n\pm}(t)\rangle\}$ with $n=0$ and $n=1$
corresponding to the ground and first excited doublets of
the double dot structure. They are pairwise degenerate at
any time  and have  energies
$\epsilon_{0,\sigma}=-\delta\epsilon(t)/2$ and
$\epsilon_{1,\sigma}=\delta\epsilon(t)/2$, with
$\delta \epsilon (t) $
the splitting between the two doublets,
\begin{equation}
\delta\epsilon(t)=\sqrt{\Delta(t)^2+v(t)^2}\;.
\end{equation}

We shall now study the electron spin's evolution within the adiabatic
approximation: we look for a solution of the Schr\"odinger
equation in the form $|\phi(t)\rangle = \sum_{n,\sigma}\alpha _{n,\sigma}(t)
|\phi_{n\sigma}(t)\rangle$. Then the wave function amplitudes satisfy the
 equation of motion
\bea
i\; \dot \alpha _{n,\sigma} &=& \sum_{n',\sigma'}
H^{\rm eff}(t)_{n\sigma,n'\sigma'}\alpha _{n',\sigma'}\;,
\\
H^{\rm eff}(t)_{n\sigma,n'\sigma'} &=&
\epsilon_n \delta_{n\sigma,n'\sigma'} + i \;\dot v \;
\langle \frac{\partial\phi_{n\sigma}}{\partial v}
|\phi_{n'\sigma'}\rangle
\nn
\\
&+&
 i \;\dot \Delta\;
\langle \frac{\partial\phi_{n\sigma}}{\partial \Delta}
|\phi_{n'\sigma'}\rangle\;.
\nn
\eea
If the time-derivatives in this expression are small compared to the splitting
$\delta\epsilon$ the evolution of the confined electron is
adiabatic and is confined to the lowest doublet of the double dot.
Remarkably, it is possible to write
down such instantaneous ground states, satisfying the
conditions $\langle \frac{\partial\phi_{0\sigma}}{\partial \Delta}
|\phi_{0\sigma'}\rangle =
\langle \frac{\partial\phi_{0\sigma}}{\partial v}
|\phi_{0\sigma'}\rangle = 0$,
\begin{eqnarray}
\phi_{0+}(\Delta,v)&\equiv &\frac{1}{\sqrt{\Delta^2+(\delta\epsilon-v)^2}}\\
&&\times\left(\delta\epsilon-v,0,-\Delta
\cos\rho ,\Delta i
e^{i\phi}\sin\rho\right)\nn\\
\phi_{0-}(\Delta,v)&\equiv &\frac{1}{\sqrt{\Delta^2+(\delta\epsilon-v)^2}}\\
&&\times\left(0,\delta\epsilon-v,\Delta i
e^{-i\phi}\sin\rho,-\Delta
\cos\rho\right)\nn
\;.
\end{eqnarray}
With this choice of basis the time evolution is trivial
in the adiabatic approximation,
and apart from an overall phase the wave function is simply given by
\be
|\phi(t)\rangle =
\alpha_- |\phi_{0-}(\Delta(t),v(t))\rangle
+ \alpha_+ |\phi_{0+}(\Delta(t),v(t))\rangle
\;.
\ee
Now imagine making an adiabatic sweep, with the potential difference
$v $ going from $v =-\infty$ at time $t=-\infty $ to $v =\infty$ at time $t=\infty $.
 The above states have been  chosen so that they
satisfy the initial condition
$|\phi_{0\pm}(-\infty)\rangle=|\phi^L_{0\pm}\rangle$ at
$t=-\infty $, and describe an electron localized in the
left potential well with
(pseudo)spins $\sigma=\pm$. According to the above
expressions, at time $t=\infty$, i.e. after the adiabatic
potential sweep $v\rightarrow\infty$, the electron will be
found fully localized in the right dot and in the following
spin superposition
\begin{equation}
|\phi(\infty)\rangle=\sum_{\sigma\sigma'}|\phi^R_{0\sigma'}\rangle
e^{-i\delta\vc r\cdot\uu\lambda_\so^{-1}\cdot{\vc
\sigma}_{\sigma'\sigma}}\langle\phi^L_{0\sigma}|\phi(-\infty)\rangle
\label{dot_precesssion}
\end{equation}
independently of $\Delta(t)$. In other words, the spin
undergoes a spin precession identical to that obtained
for $B=0$ upon adiabatically displacing a parabolic
confining potential a distance $\delta \vc r$ along a
straight line, Eq. (\ref{UadL}).   Although in our discussion we
assumed a cylindrical symmetry for the dots, a slightly modified version of
this discussion carries over to the case of non-cylindrical potentials
with  a spin rotation of comparable
size. In general, however, the spin rotation
is not given by the simple expression~\eqref{dot_precesssion}.

The most straightforward way to build a tunable gate on
this concept would be to add a magnetic flux across a ring of quantum dots as a tuning parameter. Alternatively one could use backgates on top of the heterostructure to tune the Rashba
spin-orbit coupling strengths around the ring.

\section{Conclusion}

In this article we have shown that for single electrons confined in
quantum dots in a 2DEG, which is shifted adiabatically along a path by
applied or fluctuating electric fields,
the spin-orbit interaction induces pseudospin precession
within the ground state Kramer's doublet.
In the absence of external magnetic fields, the precession depends
solely on the geometrical shape of the trajectory
of the confined electron. This accumulated
non-Abelian phase has marked
consequences for the spin relaxation and decoherence due to
electric field fluctuations. In particular it leads to a saturation of the relaxation rates at vanishing magnetic fields.
We have analyzed how the properties and power spectrum of the
electromagnetic fluctuations influences the spin relaxation rates.
We characterized two different spin decay
regimes, dominated by Ohmic or phonon-induced  electric
fluctuations, respectively, and their crossovers as a
function of external magnetic field and temperature.

The geometric spin precession analyzed in this paper
can also be used to manipulate the spin state
purely by controlling electric fields.
We have shown that arbitrary rotations can be achieved by moving
the dot along suitable trajectories. To speed-up the process we suggest
moving the electron by adiabatic tunneling between quantum dots in
multi-dot devices.

{\bf Acknowledgements:} We would like to thank W. A.
Coish and D. Loss for inspiring discussions.
This research has been supported by Hungarian grants OTKA Nos. NF061726, T046267, T046303.
G.Z acknowledges the hospitality of the CAS, Oslo.

\appendix
\section{Various matrix element relations for a parabolic potential\label{ap:Dressing}}

For the parabolic quantum dot Hamiltonian,
Eq. (\ref{H}), we have the following identity:
\begin{widetext}
\begin{eqnarray}
\langle \tau_n(t)|\left[\vco
p,\hat{H}(t)\right]|\tau'_{n'}(t)\rangle=(E_{n'\tau}-E_{n\tau})\langle
\tau_{n}(t)|\vco
p|\tau'_{n'}(t)\rangle=-im\omega_0^2\langle \tau_n(t)|\vco
r-\vc R_\mathcal{C}(t)|\tau'_{n'}(t)\rangle\nn
\end{eqnarray}
where $\vco r$ is the electron position operator, and $\vc
R_\mathcal{C}(t))$ is the dot displacement, assumed to be zero at
$t=0$. Here $E_{n\tau}$ is the eigenvalue of
the instantaneous eigenstates,
 $|\tau_n(t)\rangle$, which are related to the eigenstates of
${\mathcal H}(0)$ as  $|\tau_n(t)\rangle=\hat{\mathcal{W}}(t)|\tau_n\rangle$.
Similarly, the expectation value of $\vco r-\vc
R_\mathcal{C}(t)$ reads
\begin{eqnarray}
\left(E_{n'\tau'}-E_{n\tau}\right)\langle \tau_n(t)|\vco
r-\vc R_\mathcal{C}(t)|\tau'_{n'}(t)\rangle&=&\langle
\tau_n(t)|\left[\vco r-\vc
R_\mathcal{C}(t),\hat{H}(t)\right]|\tau'_{n'}(t)\rangle=\frac{i}{m}\langle
\tau_n(t)|\vco p+\uu\lambda_\so^{-1}\cdot \vco
\sigma|\tau'_{n'}(t)\rangle\nn
\end{eqnarray}
Combining both equations above we obtain
\begin{equation}
-\langle \tau_n|\vco
p|\tau'_{n'}\rangle=\frac{1}{1-(E_{n\tau}-E_{n'\tau'})^2/\omega_0^2}\;\uu{\lambda}_\so^{-1}\cdot
\langle \tau_n| \vco \sigma|\tau'_{n'}\rangle\;.
\end{equation}
We made use of the fact that ${\mathcal W}(t)$ commutes both with
$\vco p$ and $\vco \sigma$, and therefore
$\langle \tau_n(t)| \vco \sigma|\tau'_{n'}(t)\rangle=\langle \tau_n| \vco \sigma|\tau'_{n'}\rangle$
 and
$\langle \tau_n(t)| \vco p|\tau'_{n'}(t)\rangle = \langle \tau_n| \vco p|\tau'_{n'}\rangle$.
\end{widetext}
Specifically, for the ground state doublet (or lowest lying two states)
this equation reduces to the   useful identity
\begin{equation}
-\hat{P}_0\vco p\hat{P}_0=\uu{\tilde\lambda}_\so^{-1}\cdot
\vc\tau_{\tau\tau'}\;, \label{Plambda}
\end{equation}
where  $\vc\tau_{\tau\tau'}$
are usual Pauli matrices. All spin-dressing corrections
are contained in the matrix $\uu{\tilde\lambda}_\so^{-1}$, whose definition
trivially follows from Eq.~\eqref{Plambda}, and simplifies to
Eq.~\eqref{eq:renorm} in the absence of magnetic field.

\section{Pseudospin evolution under quantum fields \label{ap:quantumfields}}

In this appendix we generalize Eq. (\ref{UadL}) obtained for an electron
confined in a parabolic well to the case
where electromagnetic  fields are
produced by a quantum bath, which  we shall treat within the path integral
formalism. We assume that the bath is governed by Hamiltonian $\mathcal{H}_B$,
while the coupling between the bath and quantum dot is of dipolar form,
$\mathcal{V}=e\vco E\cdot\vco r$. We further allow for a Zeeman field $\vc B$
coupled to the dot,
\begin{equation}
\mathcal{H}_D
=\frac{\vco{p}^2}{2m}+V(\vco{r})+\mathcal{H}_\so+\frac{g}{2}\mu_B\vc
B\cdot\vco\sigma
\;.
\end{equation}

To combine the  path integral approach with the adiabatic approximation,
we write the evolution operator of the coupled system as
\begin{equation}
U=e^{-i(\mathcal{H}_D+\mathcal{H}_B+\mathcal{V})t}=\left[e^{-i(\mathcal{H}_D+\mathcal{H}_B+\mathcal{V})\Delta
t}\right]^{t/\Delta t}
\end{equation}
and insert the identity operator
\begin{equation}
\hat{\mathbf{1}}=\sum_{\vc E,\alpha}\vco{\mathcal{W}} |\vc E,
\alpha\rangle\hat{\mathbf{1}}_D\langle\vc E,
\alpha|\vco{\mathcal{W}}^+
\;.
\label{idW}
\end{equation}
after each time slice.
Here $\vco 1_D$ denotes the identity operator acting on the dot,
and $\vco{\mathcal{W}}\equiv \exp(-i\vco p\cdot\vco R_\mathcal{C})$.
Note that the  $\vco{\mathcal{W}} =\exp(-i\vco p\cdot\vco R_\mathcal{C})$
acts both on the bath and on the quantum dot, and now $\vco
R_\mathcal{C}=-e\vco E/m\omega_0^2$ is an operator instead
of a c-number.  In Eq.~\eqref{idW} we
also  made use of the fact that the electric field
operator $\vco E$ is Hermitian and
one  can construct a complete
basis from its eigenstates. However, $\vco E$ being a local operator,
every such state is infinitely degenerate. We keep track of this internal
degeneracy by the label $\alpha$.

The matrix elements connecting two consecutive identity
insertions labeled by $n+1$ and $n$, take the following
\begin{eqnarray}
&&\langle\vc
E_{n+1}\alpha_{n+1}|{\vco{\mathcal{W}}}^+e^{-i(\mathcal{H}_D+\mathcal{H}_B+\mathcal{V})\Delta
t}{\vco{\mathcal{W}}}|\vc E_n\alpha_n\rangle~~
\label{slice}\\
&&=\langle \vc E_{n+1}\alpha_{n+1}|e^{-i\mathcal{H}_B\Delta t}|\vc
E_n\alpha_n\rangle \times
\nn\\
&&\phantom{nnnn}
\mathcal{W}^+_{\vc
E_{n+1}}e^{-i(\mathcal{H_D}+e\vc E_n \vc r)\Delta
t}\mathcal{W}_{\vc E_n}\nn
\end{eqnarray}
where the operator $\mathcal{W}_{\vc E_n}\equiv e^{-i\vco
p\cdot \vc R_{\mathcal{C}}^n}$, $\vc R_{\mathcal{C}}^n=-e
\vc E_n/m\omega_0^2$ acts now only on the dots subspace, just
as in Sec.~\ref{subsub:parabolic}.

Using $\vc E_{n+1}\approx \vc E_{n}+\Delta t\dot{\vc E}_n$,
the second term can be expanded in $\Delta t$
and written as
\bea
&&\mathcal{W}^+_{\vc
E_{n+1}}e^{-i(\mathcal{H_D}+e\vc E_n \vc r)\Delta
t}\mathcal{W}_{\vc E_n}\nn
\nn\\
\phantom{nn}
&&
\approx 1 - i\Bigl [{\mathcal H}_D -\vco p\cdot\dot{\vc
R}^n_\mathcal{C} + \frac{e^2 \vc E_{n}^2}{2m\omega_0^2}\Bigr]
\Delta t
\nn
\eea
After re-exponentiating this expression  Eq.~(\ref{slice}) simply becomes
\begin{equation}
\approx \langle \vc E_{n+1}\alpha_{n+1}|e^{-iH'_B\Delta t}|\vc
E_n\alpha_n\rangle e^{-i(\mathcal{H_D}-\vco p\cdot\dot{\vc
R}^n_\mathcal{C})\Delta t}\label{slice2}
\end{equation}
where $H'_B=H_B-e^2\;\vc E_{n}^2/(2m\omega_0^2)$ is
effective bath Hamiltonian.

The rest of the derivation follows the standard construction of the path
integral except that we also insert the identity
operator \eqref{idW} before and after the evolution operator
$U(t)$.
The  evolution operator of the dot for   fixed  initial and final bath
states $\vc E_i$ and $\vc
E_f$, takes the form
\begin{eqnarray}
&&\langle \vc E_f|U(t)|\vc E_i\rangle=\int_{\vc E_i}^{\vc
E_f}\mathcal{D}[\vc
E]e^{-iS'_B}\label{Uquantum}\\
&&\phantom{nnnn}\mathcal{W}_{\vc E_f}{\rm T}\;\left\{e^{-i\int_0^t
dt'[\mathcal{H_D}-\vco p\cdot\dot{\vc
R}_\mathcal{C}(t')]}\right\}\mathcal{W}_{\vc E_i}^+\nn\;.
\end{eqnarray}
 Here the
functional integral is performed over all possible paths of
bath states with definite $\vc E(t)$ compatible with the
endpoints $\vc E_i$ and $\vc E_f$, each of them
corresponding to a different path $\mathcal{C}$ of the
displacement $\vc R_\mathrm{C}(t)$. Bath paths begin at
$t'=0$ and end at $t'=t$.
The weight  $e^{-iS'_B}\equiv e^{-iS'_B[E(t)]}\equiv
\int\mathcal{D}[\alpha(t)]e^{-iS'_B[E(t),\alpha(t)]}$
comes from the  prefactor in (\ref{slice2}), and
involves the effective bath Hamiltonian $H'_B$. Its
dependence on the extra quantum numbers $\alpha(t)$ is
already integrated out. Finally, the operator  $T\{e^{-i \dots}\}$ inside the
integral is acting only
on the quantum dot, and has exactly the same form as Eq.~(\ref{U})
for a classical field.

Since electric field fluctuations are assumed to be slow and the magnetic
field is small,
the time ordered operator in Eq.~\eqref{Uquantum} can be approximated by its
adiabatic form. Projecting this operator to the ground state subspace of
${\mathcal H}$ with $\vc B=0$, we have
\be
\mathcal{H_D}-\vco p\cdot\dot{\vc R}_\mathcal{C}(t')
\to
P_0\mathcal{H_D} P_0 -P_0 \vco p P_0 \cdot\dot{\vc
R}_\mathcal{C}(t') .
\ee
After using identity (\ref{Plambda}) of the previous appendix this
expression reduces to Eq.~\eqref{Uadquantum}, given in the main
text.

\section{Diagrammatic calculation of relaxation and decoherence
rates\label{ap:diagramatic}}

We can generalize the calculation of the previous appendix to compute the evolution of the
center of mass  reduced density matrix, defined as in Eq. (\ref{rhodef}),
\be
\tilde \rho_D(t)_{\tau\tau'}\equiv \langle\tau_0|\Tr_B\left[
\hat{\mathcal{W}}^+\rho(t)\hat{\mathcal{W}}\right]|\tau'_0\rangle\;.
\ee
We use the Schr\"odinger representation and insert the identity
operator \eqref{idW} for both the forward and
backward propagation in the expression of the full density matrix
$\rho(t) = e^ {-i \mathcal H\; t} \rho(0) e^ {i \mathcal H\; t}$.
This leads to the following expression [recall definitions (\ref{eq:H_Z}) and (\ref{eq:H_G})]
\begin{eqnarray}
\tilde \rho_D(t)&=&\int \mathcal{D}[\vc E] e^{-i\tilde{S}'_B}{\rm T}_K
\left[e^{-i\int_K
dz\left(H_Z+H_G\right)} \tilde \rho(0) \right]
\;,
\nn
\end{eqnarray}
where $\tilde \rho(0)= \hat {\mathcal{W}}^+\rho(0)\hat{\mathcal{W}}$ denotes the
initial center of mass density matrix.
The integration should be performed on the complex  contour,
$z=\{ 0\to t\to 0\} $, and ${\rm T}_K$ denotes the time ordering along this
contour.

We define  an operator describing the propagation of
$\tilde \rho_D$ between different times
\begin{equation}
\tilde
\rho_D(t)_{\tau_1\tau_2}=\sum_{\tau_1'\tau_2'} \Pi(t,0)_{\tau_1\tau_2\leftarrow\tau'_1\tau'_2}\tilde
\rho_D(0)_{\tau'_1\tau'_2} .
\end{equation}
In perturbation theory in the geometric coupling $H_G$ between the
dot and the electric field we can construct a corresponding Dyson equation for the
propagator $\Pi$,
$$
\Pi(t,0)=\Pi^0(t,0)+\int_0^t dt_1
dt_2\Pi^0(t,t_1)\Sigma(t_1,t_2)\Pi(t_2,0)
$$
where
$$
\Pi^0(t,0)_{\tau_1\tau_2\leftarrow\tau'_1\tau'_2}=\langle\tau_1|
e^{-itH_Z}|\tau_1'\rangle\langle\tau'_2|e^{itH_Z}|\tau_2\rangle
$$
is the bare propagator. In the following, we shall assume that the initial center of mass
density matrix factorizes as $\tilde \rho(0)=\tilde \rho_D(0) \times
\tilde\rho_{bath}(0)$, where $\tilde\rho_{bath}(0)$ represents some density
matrix of a non-interacting Gaussian heat bath.

\begin{figure}
\includegraphics[width=8 cm]{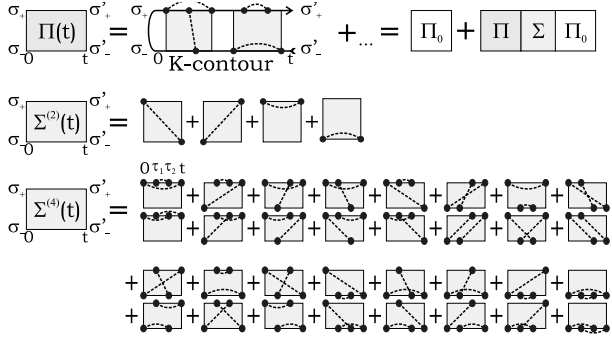}
\caption{Diagrams involved in the evaluation of the
relaxation times. The $4\times 4$ self-energy matrix
$\Sigma$ that dresses the propagator $\Pi$ is calculated to
fourth order in the dot-bath coupling (blue dots).
Contractions of bath fields in $\Sigma^{(4)}$ can be
classified in two distinct types (upper and lower rows of
diagrams). Times $\{0,\tau_1,\tau_2,t\}$ are ordered, and
$\tau_1,\tau_2$ must be integrated.\label{fig:diagrams}}
\end{figure}

Then the 'self-energy'
 $\Sigma(t)$ has an expansion in Feynman diagrams along the
$K$ contour, with vertices at  branch $s=\pm $ corresponding to terms
$-s\;i H_G$, in the expansion (see Fig. \ref{fig:diagrams}).
Differentiating with respect to $t$ and defining the Liouvillian as
$L_0 \tilde \rho_D(t)\equiv  \frac d{dt}{\Pi}_0(t)\rho_D(0)= i\left[\tilde \rho_D(t),H_Z\right]$,
one arrives at the master equation
$$
\dot{\tilde \rho}_D(t)=L_0\tilde \rho_D(t)+\int_0^t\Sigma(t-t')\tilde \rho_D(t')\;.
$$
This equation can be simplified under the Markovian approximation\cite{Makhlin02},  where we
assume that relaxation and dephasing are slow and therefore replace
$\tilde \rho_D(t')\approx e^{L_0(t'-t)}\tilde \rho_D(t)$ in the second integrand
to  yield
$$
\dot{\tilde \rho}_D(t)=L_0\tilde \rho_D(t)+\Gamma\tilde \rho_D(t)\;,
$$
with the Bloch-Redfield tensor defined as
$$\Gamma=\int_0^\infty\Sigma(t)e^{-L_0 t} \; dt\;.
$$
Relaxation and dephasing
times $T_1$ and $T_2$ are then trivially related to this tensor
$\Gamma$ as\cite{Makhlin02}
\begin{eqnarray}
T_1^{-1}&=&\Gamma_{\uparrow\uparrow\leftarrow
  \downarrow\downarrow}+\Gamma_{\downarrow\downarrow\leftarrow
  \uparrow\uparrow}\;,
\\
T_2^{-1}&=&-\mbox{Re}\;\Gamma_{\uparrow\downarrow\leftarrow
\uparrow\downarrow}\;.
\end{eqnarray}
A 'rotating wave approximation' is implicit in these relations that requires $T_1^{-1}, T_2^{-1}<\omega_B$ in our particular case, see Sec. \ref{sec:clafluc}.

Let us now perform a  calculation of $T_1$ and $T_2$
to fourth order in the coupling between the dot and the
electromagnetic field, for a small quantum dot with negligible spin dressing
($\uu{\tilde\lambda}_\so={\uu\lambda}_\so$ and $\tilde{\vc
B}=\vc B$).
Assuming an in-plane magnetic field at an angle $\theta$
with respect to the direction [100] we have
$$
H_Z=\frac{g}{2}\mu_B B\left(\cos \theta\;\hat{\sigma}_x+\sin
\theta\;\hat{\sigma}_y\right)
$$
The eigenvalues of $H_Z$ are  $\pm \omega_B/2$, where
$\omega_B\equiv-g\mu_B B$. In the particular doublet basis which
diagonalizes $H_Z$ we can write
$H_Z=-\hat{\tau}_z\omega_B/2$, and
$\Pi^0$ equals
$$
\Pi^0(t)=e^{L_0 t}=\left(\begin{array}{cccc}
1&0&0&0\\
0&e^{-i\omega_B t}&0&0\\
0&0&e^{i\omega_B t}&0\\
0&0&0&1
\end{array}\right).
$$
with the indices
ordered as
$\{\uparrow\uparrow,\uparrow\downarrow,\downarrow\uparrow,\downarrow\downarrow\}$).
This basis is rotated with respect to the original one by an
operator
$U_R=\exp\left[-\frac{i}{2}(\sin\theta\;\hat{\sigma}_x-\cos\theta\;\hat{\sigma}_y)\right]$,
so that our vertex  $-isH_G$ in branch $s=\pm$, reads in this
basis
\begin{eqnarray*}
-is H_G&=&-is\sum_{\mu,\nu=x,y}\dot{
R}_{\mathcal{C}\mu}(t_s){\lambda^{-1}_{\so\mu\nu}}U_R\hat{\sigma}_\nu
U_R^+\\
&=&-is
\mathop{\sum_{\mu=x,y}}_{\nu=x,y,z}\vc{\gamma}^\nu_\mu{\hat\tau}_\nu\frac{\dot{R}_{\mathcal{C}\mu}(t_s)}{x_0}
\end{eqnarray*}
where
$\vc{\gamma}_\mu=\{\gamma^x_\mu,\gamma^y_\mu,\gamma^z_\mu\}=\{-\sin\theta\gamma_\mu^\perp,\cos\theta\gamma_\mu^\perp,\gamma_\mu^\parallel\}$
and the relevant transverse and parallel dimensionless
couplings, $\gamma_\mu^\perp$ and $\gamma_\mu^\perp$,
are defined by
\begin{equation}
\left(\begin{array}{cc}
\gamma^\perp_x&\gamma^\perp_y\\
\gamma^\parallel_x&\gamma^\parallel_y
\end{array}\right)=
m x_0\left(\begin{array}{cc}
-\alpha\cos\theta+\beta\sin\theta&\beta\cos\theta-\alpha\sin\theta\\
-\beta\cos\theta-\alpha\sin\theta&\alpha\cos\theta+\alpha\sin\theta
\end{array}\right)
\end{equation}
For compactness,  we shall use the vector notation
 $\vc
{\gamma}^{\perp,\parallel}\equiv\{\gamma^{\perp,\parallel}_x,\gamma^{\perp,\parallel}_y\}$
in the rest of this appendix.
The corresponding vertex matrices
$V_+(t)=-iH_G\otimes\mathbf{1}$ and
$V_-(t)=i\mathbf{1}\otimes H_G^T$ at time $t$ at branch $s=\pm$
that enter the expansion of $\Sigma(t)$
(are denoted by blue dots in Fig.
\ref{fig:diagrams}, and  read
\begin{eqnarray}
V_+(t)=\sum_\mu \left(\begin{array}{cccc}
-i \gamma^\parallel_\mu&0&e^{i\theta}\gamma^\perp_\mu&0\\
0&-i\gamma^\parallel_\mu&0&e^{i\theta}\gamma^\perp_\mu\\
-e^{-i\theta}\gamma^\perp_\mu&0&i\gamma^\parallel_\mu&0\\
0&-e^{-i\theta}\gamma^\perp_\mu&0&i\gamma^\parallel_\mu
\end{array}\right)\dot{R}_{\mathcal{C}\mu}(t_+)\nn\\
V_-(t)=\sum_\mu \left(\begin{array}{cccc}
i \gamma^\parallel_\mu&e^{-i\theta}\gamma^\perp_\mu&0&0\\
-e^{i\theta}\gamma^\perp_\mu&-i\gamma^\parallel_\mu&0&0\\
0&0&i\gamma^\parallel_\mu&e^{-i\theta}\gamma^\perp_\mu\\
0&0&-e^{i\theta}\gamma^\perp_\mu&-i\gamma^\parallel_\mu
\end{array}\right)\dot{R}_{\mathcal{C}\mu}(t_-)\nn
\end{eqnarray}
 In this notation we have the following relation for
the $n$-th order   contribution to  Bloch-Redfield tensor $\Gamma$
in the dot-bath
coupling $V(t)\equiv V_+(t)+V_-(t)$
\bea
\Gamma^{(n)}&=&\int_{0}^\infty dt_2\dots dt_n
\\
&&\langle
V(t_n) \Pi^0(t_n-t_{n-1}) \dots
V(0)\Pi^0(0-t_n)\rangle_B\, .
\nn
\eea
Here $t_n > \dots >t_2$, i.e. a time-ordered integral over the internal times
is implicit, and the bracket
$\langle\dots\rangle_B$ denotes averaging over
the bath fields
$\dot{R}_{\mathcal{C}\mu}$, i.e. pairwise contractions for
a non-interacting (Gaussian) bath. Only connected contributions must be taken
into account in the averaging, and all odd $n$ contributions average to
zero. The $n=2$ and $n=4$ diagrams are represented in
Fig. \ref{fig:diagrams}. The $n=4$ case has two types of
distinct contractions that correspond to contracting the vertices at
$t_1=0$ and $t_4$ with each-other and with internal vertices, respectively
(represented in two separate
rows in Fig. \ref{fig:diagrams}). Each contraction
gives  a non-interacting bath
Green's function  defined in the Heisenberg picture as
\begin{eqnarray}
G_{\mu\nu}^{s'\!\!,s}(t'-t)&=&-i\frac{1}{x_0^2}\langle
\dot{R}_{\mathcal{C}\mu}(t'_{s'})\dot{R}_{\mathcal{C}\nu}(t_{s})\rangle_B\nn\\
&=&-i\frac{1}{x_0^2}\Tr_B\left[T_{K}\dot{\hat R}_{\mathcal{C}\mu}(t'_{s'})\dot{\hat R}_{\mathcal{C}\nu}(t_s)\rho_B(0)\right]\nn\\
&=&-i\frac{1}{x_0^2}\int \mathcal{D}[\vc E]
e^{-i\tilde{S'}_B}\dot{R}_{\mathcal{C}\mu}(t'_{s'})\dot{R}_{\mathcal{C}\nu}(t_s)\nn
\;.
\end{eqnarray}
Here once more $s,s'=\pm$ refer to the time branch,
and  the time evolution of the operators $\dot{\hat
R}_{\mathcal{C}\mu}(t)$ in the first line is governed by $H'_B$.
For simplicity, we assume an isotropic bath so that
$G_{\mu\nu}^{s'\!\!,s}=\delta_{\mu\nu}G^{s'\!\!,s}$.

In the literature, one
commonly denotes in the off-diagonal components as  $G^<\equiv
G^{+-}$ and $G^>\equiv G^{-+}$. The spectral function is
related to these as $A\equiv i(G^>-G^<)$,
while the Keldysh
propagator is given by $G^K\equiv G^>+G^< $. Since $E(t)$ is a real
bosonic field, in equilibrium we have
\bea
G^>(\omega)&=&G^<(-\omega)=-i\left(1+n_B(\omega)\right)A_{\vc{\dot{R}}}(\omega),\nn\\
G^K(\omega)&=&-i \coth\left(\frac{\omega}{2k_B T}\right)A_{\vc{\dot{R}}}(\omega),
\label{eq:spectral_connection}
\eea
where $n_B(\omega)$ is the Bose distribution, and
$A_{\vc{\dot{R}}}(\omega)$ is the spectral function of the
rescaled bath field $\vc{\dot{R}}/x_0$. This spectral function
can be easily related to the spectral function $A(\omega)$ of the
dimensionless electric field, $ex_0\vc E/\omega_0$ as
$A_{\vc{\dot{R}}}(\omega)=\omega^2A(\omega)$. Both spectral functions are odd in $\omega$.

\begin{figure}
\includegraphics[width=8 cm]{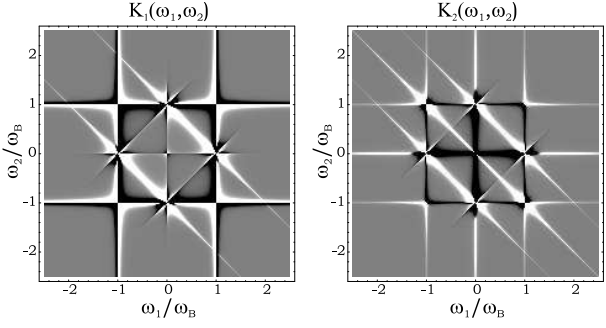}
\caption{The kernels $K_1$ and $K_2$ involved in the result
for $T^{(4)}_{1,2}$, Eq. (\ref{kernels}) for generic values
of the coupling constants $\vc\gamma^{\perp,\parallel}$. A
resolving imaginary part $\epsilon=0.01\omega_B$ was used
to make the delta functions visible.\label{fig:kernels}}
\end{figure}

The second order results for $T_1$ and $T_2$ take the known form
\cite{Leggett87,Weiss99}
\begin{eqnarray}
{T_1^{(2)}}^{-1}&=& 2|\vc\gamma^\perp|^2iG^K(\omega_B)\\
{T_2^{(2)}}^{-1}&=&\frac{1}{2}{T_1^{(2)}}^{-1}+|\vc\gamma^\parallel|^2iG^K(0)
\end{eqnarray}
The fourth order contributions to $\Gamma$ are obtained by summing all
corresponding diagrams. They are rather involved, but have the
general structure
\begin{equation}
{T_{1,2}^{(4)}}^{-1}=\int_{-\infty}^\infty d\omega_1
d\omega_2
K_{1,2}(\omega_1,\omega_2)iG^>(\omega_1)iG^>(\omega_2).\label{kernels}
\end{equation}
 The kernel
$K_\kappa$ contains  delta
functions of $\omega_1,\omega_2$. In Fig. \ref{fig:kernels}
we plot the lines along which these delta functions
pick up their contribution for  relaxation and
dephasing. We can distinguish  two types of  lines. The diagonal lines
in the 2nd and 4th quadrants lead to convergent integrals,
since $G^>(\omega)G^>(-\omega)$ goes exponentially to zero
at large values of $\omega$. Diagonal lines with positive
slope and also horizontal and vertical lines across the
origin cancel out exactly in $K_1$. The remaining horizontal and
vertical lines exhibit ultraviolet divergencies, and
give therefore cutoff dependent prefactors that multiply
$G^K(\pm\omega)$ and $G^K(0)$. These terms can thus be
reabsorbed  in the  second order
result by simply renormalizing the couplings $\vc\gamma^{\perp,\parallel}$.
The final result can then be expressed in terms of these renormalized
constants as
\begin{eqnarray}
T_1^{-1}&=&2|\vc\gamma^\perp|^2iG^K(\omega_B)\\
&&+2|\vc\gamma^\perp|^2|\vc\gamma^\parallel|^2
iF^K_1(\omega_B)+2\left(\vc\gamma^\perp\cdot\vc\gamma^\parallel\right)^2iF^K_2(\omega_B)\nn\\
T_2^{-1}&=&\frac{1}{2}{T_1}^{-1}+|\vc\gamma^\parallel|^2iG^K(0)+|\vc\gamma^\perp|^4iF(\omega_B)
\end{eqnarray}
where the $F$ functions are given by the following
convolutions of $G^>(\omega)$
\begin{eqnarray}
F^K_{1,2}(\omega)&=&F_{1,2}^>(\omega)+F_{1,2}^>(-\omega)
\;,
\nn\\
F^>_1(\omega)&=&\int_{-\infty}^\infty\frac{d\omega'}{2\pi}iG^>\left(\frac{\omega}{2}-\omega'\right)
iG^>\left(\frac{\omega}{2}+\omega'\right)\nn\\
&&\mbox{Re}\left[\left(\frac{1}{\frac{\omega}{2}-\omega'-i0}\right)^2+\left(\frac{1}{\frac{\omega}{2}+\omega'+i0}\right)^2\right]
\;,
\nn\\
F^>_2(\omega)&=&\int_{-\infty}^\infty\frac{d\omega'}{2\pi}iG^>\left(\frac{\omega}{2}-\omega'\right)
iG^>\left(\frac{\omega}{2}+\omega'\right)\nn\\
&&\mbox{Re}\left[2\frac{1}{\frac{\omega}{2}-\omega'-i0}\frac{1}{\frac{\omega}{2}+\omega'+i0}\right]
\;,
\nn\\
F(\omega)&=&\int_{-\infty}^\infty\frac{d\omega'}{2\pi}iG^>\left(-\omega'\right)
iG^>\left(\omega'\right)
\nn\\
&&\mbox{Re}\left[\left(\frac{1}{\omega-\omega'-i0}+\frac{1}{\omega+\omega'+i0}\right)^2\right]
\;.
\label{Fs}
\end{eqnarray}
The Green's functions $G^<$ and $G^>$ can be expressed in terms of
the spectral function $A(\omega)$  of the dimensionless electric field
$ex_0\vc E/\omega_0$, introduced below Eq.~\eqref{eq:spectral_connection}
\begin{eqnarray}
iG^>(\omega)&=&\left(1+n_B(\omega)\right)\omega^2
A(\omega)
\;,
\nn\\
iG^K(\omega)&=&\coth\left(\frac{\omega}{2k_BT}\right)\omega^2A(\omega)
\;.
\end{eqnarray}
Inserting these last equations into the expression
\eqref{Fs} one arrives at the results in Eqs.~(\ref{T1f})
and (\ref{T2f}). The spectral function $A(\omega)$ for electromagnetic fluctuations
generated by piezoelectric phonons  is computed in the next appendix.

\section{Phonon bath properties \label{ap:phonons}}

Let us consider the fluctuating electric field induced by
the phonons in the sample holding the quantum dot. The
electric field acting on the confined electron is the gradient of the potential generated
by these phonons, $e E_\mu=-\nabla_\mu
U^{\mathrm{ph}}(\hat{x},\hat{y})$. It has two
contributions, one from a longitudinal mode and another
from two transverse modes
\begin{equation}
U^{\mathrm{ph}}=\frac{1}{\sqrt{V}}\sum_{\vc{q},\lambda}e^{i\vc{r}\vc{q}}M_{\vc{q},\lambda}b^+_{\vc{q},\lambda}+\mathrm{h.c.}
\end{equation}
with $\lambda=l,t_1,t_2$ indicating the mode. The
coupling to the longitudinal is given by \cite{Cheng04}
\begin{eqnarray}
M_{\vc{q},l}^2&=&\frac{e^2h_{14}^2}{2\rho v_l
q}\left(\frac{3 q_xq_yq_z}{q^3}\right)^2J(q_z d_w)\nn\;,
\end{eqnarray}
while the coupling to the transverse mode is
\begin{eqnarray}
M_{\vc{q},t_1}^2+M_{\vc{q},t_2}^2&=&\frac{
e^2h_{14}^2}{2\rho v_t
q}J(q_z d_w)\nn\\
&&\frac{q_x^2q_y^2+q_y^2q_z^2+q_z^2q_x^2-9
q_x^2q_y^2q_z^2/q^2}{q^5}\nn\;.
\end{eqnarray}
Here $q\equiv |\vc{q}|$, $V$ is volume, $d_w$ is the
depth of the quantum well where the 2-dimensional electron gas is confined.  The function
$J(x)=\Theta(1-x^2)$, $\Theta$ being the Heaviside
function, qualitatively accounts for the truncation of the
phonon spectrum out of the 2DEG plane.\cite{Golovach04}
The physical origin of this cut-off is that phonons having a
wave-vector component larger  than $\sim 1/d_w$ along the $z$ direction
cannot couple  efficiently  to the confined electron, since their wave
function  oscillates too quickly.  The density $\varrho$, the  sound velocities $v_l$ and $v_t$, and
the piezoelectric constant $h_{14}$ in the expressions above
are material-dependent parameters, which
depend on the particular heterostructure used to define the quantum dot:
for a typical GaAs/AlGaAs
heterostructures  $\rho=5.3\cdot 10^3 \;\mathrm{Kg/m^3}$,
 $v_l=4.73\cdot 10^3\;\mathrm{m/s}$, $v_t=3.35\cdot 10^3\;\mathrm{m/s}$, and
$h_{14}=1.4 \cdot 10^9\;\mathrm{V/m}$.

The noise power $S^>(t'-t)=\langle
E_\mu(t'_-)E_\mu(t_+)\rangle_Be^2x_0^2/\omega_0^2$ of the
normalized electric field can be expressed as
\bea
S^>(t)&=&\frac{x_0^2}{V\omega_0^2}\sum_{\vc{q},\lambda}q_i^2M_{\vc{q},\lambda}^2
\nn\\
&&\left(\langle
b^+_{\vc{q},\lambda}(t)b_{\vc{q},\lambda}(0)\rangle+\langle
b_{\vc{q},\lambda}(t)b^+_{\vc{q},\lambda}(0)\rangle\right)\;.\nn
\eea
It can be checked that $E_x$ and $E_y$ are indeed
independent in this model.

Using $\langle
b_{\vc{q}}(t)b^+_{\vc{q}}(0)\rangle=e^{-i\omega_q
t}[1+n_B(\omega_q)]$, $\langle
b^+_{\vc{q}}(t)b_{\vc{q}}(0)\rangle=e^{i\omega_q
t}n_B(\omega_q)$ and $\omega_q=v_l q$ we see that the only
dependence on the orientation of the wave vector vector $\vc q$ appears through $M$. We
can therefore introduce spherical coordinates in the continuum limit and integrate with
respect to the angular variables.  For the longitudinal phonons this yields
\begin{eqnarray}
S^{>}_{l}(t)&=&\frac{x_0^2 e^2h_{14}^2}{\omega_0^2\rho
v_l}\frac{3}{210\pi}\int_0^\infty \frac{dq}{2\pi}
j_l\left(qd_w\right) q^3\label{Sph}\\
&&\times\left(e^{i\omega_{q,l}
t}n_B(\omega_{q,l})+e^{-i\omega_{q,l}
t}[1+n_B(\omega_{q,l})]\right)\nn\;,
\end{eqnarray}
where the function $j_l(qd_w)$ above comes from the
truncation of the phonon spectrum, and is given by
\begin{eqnarray}
&&j_l(x>1)=\frac{-35+135x^2-189x^4+105x^6}{16 x^9}\nn
\;,
\\
&&j_l(0<x<1)=1
\;.
\end{eqnarray}

The two transverse modes give a similar contribution
\begin{eqnarray}
S^{>}_{t}(t)&=&\frac{x_0^2 e^2h_{14}^2}{\omega_0^2\rho
v_t}\frac{4}{210\pi}\int_0^\infty \frac{dq}{2\pi}
j_t\left(qd_w\right) q^3\\
&&\times\left(e^{i\omega_{q,t}
t}n_B(\omega_{q,t})+e^{-i\omega_{q,t}
t}[1+n_B(\omega_{q,t})]\right)\nn\;,
\end{eqnarray}
with
\begin{eqnarray}
&&j_t(x>1)=\frac{105-300x^2+294x^4-140x^6+105x^8}{64 x^9}\nn\\
&&j_t(0<x<1)=1
\end{eqnarray}
The cut-off functions $j_l(x)$ and $j_t(x)$ play a role
similar to $\Theta(1-x^2)$, but have
algebraic tails: $j_l(x)\sim x^{-3}$ for large $x$, while $j_t(x)\sim 1/x$.
The phonon spectrum is defined as usual by
$\omega_{q,\lambda}=v_\lambda q$.

Taking the Fourier transform of the correlation functions above,
we can identify the spectral function $A_\mathrm{ph}(\omega)$,
\begin{eqnarray}
S^>(\omega)&=&\left[1+n_B(\omega)\right]A_\mathrm{ph}(\omega)\\
A_\mathrm{ph}(\omega)&=&\frac{x_0^2}{\omega_0^2}\lambda_{\mathrm{ph}}\omega^3\\
\lambda_{\mathrm{ph}}&=&\frac{e^2h_{14}^2}{210\pi\rho
}\left[\frac{3}{v_l^5}j_l\left(\frac{\omega}{\omega_{w,l}}\right)+\frac{4}{v_t^5}j_t\left(\frac{\omega}{\omega_{w,t}}\right)\right]\nn
\end{eqnarray}
where $\omega_{w,l}\equiv v_l/d_w$ and $\omega_{w,t}\equiv
v_t/d_w$.
For  frequencies smaller than the cutoffs $\omega_{w,l}$
we obtain $\lambda_{\mathrm{ph}}=2.5\cdot
10^{-5}\mathrm{K^{-2}nm^{-2}}$ for GaAs/AlGaAs quantum wells. Note that
in this calculation we neglected the lateral size of the dot.
That provides an additional cut-off for the in-plane phonon momentum
for larger values of the frequency $\omega$.

\bibliographystyle{apsrev}
\bibliography{Berry}
\end{document}